\documentclass[12pt]{article}
\usepackage{epsfig}
\usepackage{axodraw}
 
\newcommand{\mrm}[1]{\mathrm{#1}}
 
\newlength{\abstwidth}
\begin{document}
 
%set sloppy attitude to line breaks
\sloppy
 
\pagestyle{empty}
 
\begin{flushright}
LU TP 99--39 \\
December 10, 1999
\end{flushright}
 
\vspace{\fill}
 
\begin{center}
{\LARGE\bf Effects of different Form-factors}\\[3mm]
{\LARGE\bf in Meson-Photon-Photon Transitions}\\[3mm]
{\LARGE\bf and the Muon Anomalous Magnetic Moment\footnote{Master of Science
Thesis by Fredrik Persson with thesis advisor: Johan Bijnens.
A more paper like version with more references will follow.}}\\[2cm]
{\bf Johan Bijnens and Fredrik Persson}\\[3mm]
{\it Department of Theoretical Physics, Lund University,\\[1mm]
S\"olvegatan 14A, S22362 Lund, Sweden}
\end{center}
 
\vspace{\fill}
\begin{center}
{\bf Abstract}\\[2ex]
\begin{minipage}{\abstwidth}
The exact form of the form-factor associated with a meson-photon-photon
vertex is not known. Different suggestions exist, based on constraints 
from QCD and on recent experiments. Four different form-factors are 
studied in this article. 

We calculate decay rates for the decays 
\mbox{$\pi^{0}\rightarrow \gamma\gamma$}, 
\mbox{$\eta\rightarrow \gamma\  e^{+}e^{-}$},
\mbox{$\eta\rightarrow \mu^{+}\mu^{-}\ e^{+}e^{-}$} and 
\mbox{$\eta\rightarrow e^{+}e^{-}\ e^{+}e^{-}$} and cross sections for 
the processes $e^{+}e^{-}\rightarrow PS\ e^{+}e^{-}$, where 
$PS=\pi^{0}$, $\eta$ or $\eta'$. The results depend
on the choice of form-factor and we examine if the differences are large
enough to distinguish in upcoming experiments. 

The dominant part of the light-by-light contribution to the muon 
anomalous magnetic moment is also calculated. Here the uncertainty 
around the choice of form-factor implies that it is possible to 
estimate a smallest error in the theoretical value of this contribution.

\end{minipage}
\end{center}

\vspace{\fill}
 
\clearpage

\pagestyle{plain}
\setcounter{page}{1}
                                                                        
%***********************************************************************

% SECTION (Introduction)

\section{Introduction}
\label{sec-intro}

%***********************************************************************

Understanding the structure of the natural universe can be seen as a 
three-part problem. Identifying the basic particles that are the 
constituents of matter, knowing what forces they feel and knowing how 
to calculate the behaviour of the particles given the forces. 
The Standard Model is the theory that tries to answer these questions 
today.\\ 

The particles can be put into two categories, matter particles and gauge
 bosons. The matter particles, the quarks and the leptons, are the 
building blocks of nature. Both are spin-$1/2$ fermions but the quarks 
also have colour charge as described in QCD. So far there are known to 
exist six kinds of quarks and six kinds of leptons.

The quarks are called up, down, strange, charm, top and bottom. They 
are denoted by their first letter and are divided into three doublets, 
called generations:
%
%***************Table of quarks.****************************************
\begin{displaymath}
\begin{array}{ccc}
\left( \begin{array}{c} \mrm{u} \\ \mrm{d} \end{array} \right) 
& \left( \begin{array}{c} \mrm{c} \\ \mrm{s} \end{array} \right) 
& \left( \begin{array}{c} \mrm{t} \\ \mrm{b} \end{array} \right). 
\end{array}
\end{displaymath}
%***********************************************************************
%
The top row has electric charge $q=+(2/3)e$ and the bottom row has 
$q=-(1/3)e$, where $e$ is the magnitude of the electron's charge.

The six leptons are also arranged in three generations:
%
%***************Table of leptons.***************************************
\begin{displaymath}
\begin{array}{ccc}
\left( \begin{array}{c} \nu_{\mrm{e}} \\ \mrm{e} \end{array} \right) 
& \left( \begin{array}{c} \nu_{\mu} \\ \mu \end{array} \right) 
& \left( \begin{array}{c} \nu_{\tau} \\ \tau \end{array} \right). 
\end{array}
\end{displaymath}
%***********************************************************************
%
The electron, muon and tau have electric charge $-e$ and each has its 
own neutrino of electric charge zero. The quarks and the leptons are 
the basic particles of matter.\\

The second type of particles are the gauge bosons, which all have 
integer spin. They are best described together with the forces. 
Classically, a force is seen as a field, e.g.\ a charged object is 
surrounded by an electric field. But in Quantum Field Theory, which is 
based on special relativity and quantum mechanics, an important part 
is that particles and fields are treated in the same way. Whether they 
behave as particles or fields depend on the circumstances and the 
gauge bosons are just those particles that correspond to the forces 
and their fields. There are four forces in nature and each has its own 
gauge bosons:
%
%****************Table of forces.***************************************
\begin{displaymath}
\begin{array}{lll}
\underline{\mrm{Force}} & \underline{\mrm{Acts\ on}}& 
\underline{\mrm{Gauge\ bosons}} \\
\mrm{gravitation} & \mrm{all\ particles} & \mrm{graviton}\\
\\
\mrm{electromagnetic} & \mrm{all\ electrically} & \mrm{\gamma}\\
\mrm{} & \mrm{charged\ particles} & \mrm{}\\
\\
\mrm{weak} & \mrm{quarks,leptons} & \mrm{W^{+},\ W^{-},\ Z^{0}}\\
\mrm{} & \mrm{electroweak\ gauge\ bosons} & \mrm{}\\
\\
\mrm{strong} & \mrm{all\ coloured\ particles} & 
\mrm{g}_{i},\ i=1 \ldots 8
\end{array}
\end{displaymath}
%***********************************************************************
% 
The gravitational force acts upon all particles but the strength depends
 on the mass and it can be disregarded in all but the highest 
energy calculations in particle physics. The electromagnetic force 
affects charged objects and holds e.g.\ the electrons and the nuclei 
together in atoms. It is transmitted by the photons. The weak 
interaction, or force, is needed to explain e.g.\ the 
$\beta$-decay of nuclei and it is transmitted by three very heavy 
bosons, which makes it a short-range interaction according to 
Heisenbergs uncertainty relation. The strong force holds the quarks 
together inside the nucleons as well as the protons and neutrons inside 
nuclei. The carriers are the eight gluons which themselves have 
colour charge.

Furthermore, all particles, fermions and bosons, have antiparticles. 
They have opposite values of all charges, but the same mass.
\cite{ref:Kvarken, ref:Kane}\\

Particles with colour charge like the quarks and gluons can't exist on 
their own. They are permanently bound inside colourless hadrons. 
Hadrons can be baryons, in which case they are made up of three quarks, 
or mesons, if they are made up of one quark and one anti-quark. 
Depending on how they behave under parity-transformations and rotations,
 the mesons 
can be scalar-mesons, vector-mesons etc. We study properties of the 
three pseudo-scalar mesons $\pi^{0}$, $\eta$ and $\eta'$ in this 
article. 

%***********************************************************************
\subsection{Form-factors}   
\label{subsec-Form}
%***********************************************************************

An interesting thing to keep in mind is that the known universe is made 
up of $\mrm{u}$-quarks, $\mrm{d}$-quarks, electrons, $\nu_{\mrm{e}}$ and
 the gauge bosons. All the other quarks and leptons existed at an early 
stage of the universe and can today only be seen in cosmic rays and 
accelerator experiments. Such experiments makes it possible to study 
the properties of the different interactions, i.e. how the particles 
behave given the forces.

For example, studying $\gamma\gamma$ scattering provides a unique 
opportunity to learn the properties of strong interactions. Although 
in $\gamma\gamma$ scattering the probe and the target are both photons 
that are carriers of the electromagnetic force, they can produce a 
pair of quarks that interact strongly and are observed in the form of 
hadrons, e.g.\ pseudo-scalar mesons.  However, the transition between 
a meson and two photons can't be calculated from QCD directly. 
Therefore, a temporary solution is to calculate the process without 
including all effects of the quarks and gluons and at the end take the 
neglected effects into account with an extra factor. This factor is, 
for historical reasons, known as a form-factor. The shape of this 
form-factor is however not known exactly. There exist different 
educated guesses which fits certain demands from QCD, but since one 
doesn't know exactly which one to use it is important to know 
which effects a specific choice will have. This article studies the 
variation of some physical quantities with the choice of form-factor.\\

The article is organized as follows. In section 2 the form-factors are 
presented in more detail. Section 3 includes a comparison between the 
decay-rates for different form-factors for processes involving 
$\eta$. Section 4 treats the cross-section of 
$\mrm{e}^{+}\mrm{e}^{-}\rightarrow\mrm{e}^{+}\mrm{e}^{-}PS$, 
where $PS$ is $\pi^{0}$, $\eta$ or $\eta'$, and how it is affected by 
the choice of form-factor. Section 5 contains a discussion of the 
effect the form-factors have on the theoretic value of the muon 
anomalous magnetic moment and section 6 is a summary of this article. 

\vspace{\fill}
 
\clearpage

\pagestyle{plain}

%***********************************************************************

% SECTION (Form-Factors in Meson-Photon-Photon Transitions)

\section{Form-factors in Meson-Photon-Photon Transitions}
\label{sec-Form}

%***********************************************************************

The transition between a meson and 
two photons is very hard to calculate in QCD. That depends on the fact 
that for low energies $\alpha_{s}$, the strong coupling constant, is 
too large for a perturbative approach to work. Beside perturbative 
calculations no good method exist in this case so the problem gets 
complicated and approximations have to be made.

In the approximation that the quarks are massless, QCD has chiral 
symmetry. That means that all quarks, right-handed as well as 
left-handed, are treated in the same way and give the same 
results\footnote{In reality, this symmetry is spontaneously 
broken\cite{ref:Dynamics}.}. 
According to Noether's theorem, which states;\\
\emph{For a system described by a Lagrangian, any continuous symmetry 
which leaves invariant the action, $\int\mathcal{L}\ \mrm{dt}$, 
leads to the existence of a conserved current},\\
this implies that the current, $S^\mu$, which corresponds to 
the chiral symmetry, fulfills the relation
\begin{displaymath}
\partial_{\mu}{S^{\mu}}=0.
\end{displaymath}
But when one combines QCD and electromagnetism, as in the vertex we are 
studying, this is no longer true. The four-divergence of the current 
isn't zero anymore but instead it has the form
\begin{displaymath}
\partial_{\mu}{S^{\mu}}=C\ e^{2}\ \varepsilon^{\mu\nu\alpha\beta}\ 
F^{\gamma}_{\mu\nu}\ F^{\gamma}_{\alpha\beta},
\end{displaymath}
where $C$ is a known constant, $e$ the magnitude of the electron's 
charge, $\varepsilon^{\mu\nu\alpha\beta}$ the totally antisymmetric 
Levi-Civita tensor and $F^{\gamma}$ the electromagnetic field 
strength.

This was very surprising to the people that first calculated it so it 
was given the name `the anomaly'. But since this effect exists the 
effective Lagrangian that should describe this vertex should reproduce 
it. That 
puts strong constraints on how the Lagrangian can look and there is 
actually only one way to construct it so that it fulfills this anomaly 
condition. From this Lagrangian, the term which describes the 
low-energy processes that include $\pi^{0}$, $\eta$ or $\eta'$ and two
photons can be written
\begin{displaymath}
\mathcal{L}=-\frac{e^{2}}{4\sqrt{2}\pi^{2}F_{\pi}}\ \partial_{\mu}
{A_{\nu}(x)}\ A_{\alpha}(x)\ \partial_{\beta}{\phi(x)}\ \varepsilon^
{\mu\nu\alpha\beta},				
\end{displaymath}
where $F_{\pi}$ is the pion decay constant, % \cite{ref:pion}, 
$A_{\nu}(x)$ and $A_{\alpha}(x)$ the photon fields and $\phi(x)$ the 
pseudo-scalar field.

\begin{figure}[hb]
\begin{center}
\begin{picture}(300,90)
\Vertex(150,45){2}
\Line(150,45)(200,45)
\put (210,42){$\pi^{0}$, $\eta$, $\eta'$}
\Photon(70,85)(150,45){3}{4.5}
\Photon(70,5)(150,45){-3}{4.5}
\put(55,82){$\gamma$}
\put(55,2){$\gamma$}
\end{picture}
\end{center}
\caption{Feynman diagram for the transition when the reaction is seen 
as pointlike.}
\label{fig:pointlike}
\end{figure}
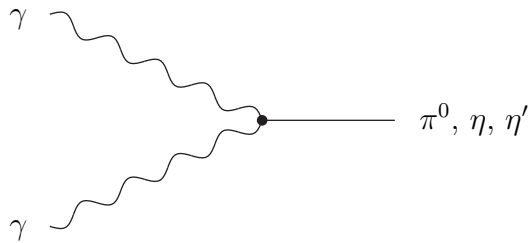

However, this Lagrangian only describes the process in the limit where 
the pseudo-scalar is seen as a point-like particle, with no inner 
structure, and the reaction is seen as in Fig.~\ref{fig:pointlike}.
%
%***********************Figure of Pointlike Reaction********************
%***********************************************************************
%
%
%*********************Figure of More Realistic Reaction*****************
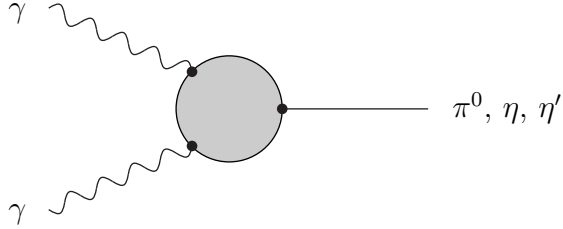
\begin{figure}[t]
\begin{center}
\begin{picture}(300,100)
\GCirc(150,50){20}{0.8}
\Vertex(170,50){2}
\Vertex(136,64){2}
\Vertex(136,36){2}
\Photon(82,88)(136,64){3}{4.5}
\Photon(82,12)(136,36){-3}{4.5}
\Line(170,50)(225,50)
\put(235,47){$\pi^{0}$, $\eta$, $\eta'$}
\put(67,85){$\gamma$}
\put(67,9){$\gamma$}
\end{picture}
\end{center}
\caption{The realistic situation with more quark and 
gluon effects.}
\label{fig:shaded}
\end{figure} 
%***********************************************************************
%
In reality, these particles are not pointlike but instead made up of 
quarks and gluons and the reaction is more like in 
Fig.~\ref{fig:shaded},
 where the shaded area includes different effects from quarks and 
gluons. If these effects are taken into account other terms come in 
which are added to the point-like Lagrangian. The new terms all have 
to fulfill the condition $\partial_{\mu}{S^{\mu}}=0$, which means that 
they don't add anything to the four-divergence and the total 
Lagrangian still meets the requirement of the anomaly condition. But 
these new quark and gluon terms are very difficult to handle so they 
are omitted and their effects are only introduced as a form-factor. 
That is, a factor is multiplied to the vertex which takes into 
account all the effects that the extra terms would add. But since 
these effects are not completely known, the shape of the form-factor 
is uncertain. Still, there exist some constraints on it.

%***********************************************************************
\subsection{Constraints on the Form-factors}   
\label{subsec-Constraints}
%***********************************************************************

The form-factor connects three particles and therefore depends on three
variables, the momentum-squared of the photons and the pseudo scalar. 
However, since we work under the approximation that the quark-mass is
zero, the pseudo scalar will also become massless. Unlike the photons, 
which can be virtual, this will mean that $q^{2}_{PS}$ will always
be zero and we set
\begin{displaymath}
F(k^{2}_{1},k^{2}_{2},0)=F(k^{2}_{1},k^{2}_{2}),
\end{displaymath}
where $k_{1}$ and $k_{2}$ are the four-momenta of the two photons.\\
 
It would be very hard to find a form-factor that describes this 
transition well if one didn't have any constraints on how it should 
behave. Luckily, it is possible to calculate, directly from QCD, how it
looks in certain limits. There are actually four different requirements
on the shape of the form-factor. Three which are demands from 
QCD-calculations and one of a more argumentative nature. 

The first limit in which the form-factor is known is when 
$k^{2}_{1}=k^{2}_{2}=0$, that is, when the two 
photons are real. Then the transition can be compared to 
other processes like $\pi^{+}\rightarrow\mu^{+}\ \nu_{\mu}$, and it is 
actually in this limit that the Lagrangian is defined and $F_{\pi}$ 
is set. So here, by definition, all quark and gluon effects are 
included in the pointlike Lagrangian and the form-factor must 
fulfill the relation
\begin{displaymath}
F(0,0)=1.
\end{displaymath}
The next case where it is possible to calculate the result from QCD is
when both photons are very much off shell and have the same virtuality,
 that is $k_{1}^{2}=k_{2}^{2}<<0$. In this case one can use what is 
called the operator product expansion technique and calculate the 
matrix-element related to this transition up to order $1/k^{2}$.  
This gives the answer
\begin{displaymath}
F(k^{2}_{1},k^{2}_{2})=-\frac{8\pi^{2}\ F^{2}_{\pi}}{N_{c}\ k^{2}},
\ \ \ \ \ k_{1}^{2}=k_{2}^{2}=k^{2}<<0,
\end{displaymath} 
where $N_{c}$ is the number of colours in the Standard Model, which
is three.

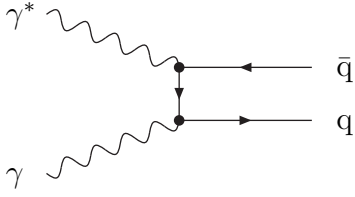
\begin{figure}
\begin{center}
\begin{picture}(300,100)
\Vertex(150,40){2}
\Vertex(150,60){2}
\ArrowLine(150,40)(200,40)
\ArrowLine(200,60)(150,60)
\ArrowLine(150,60)(150,40)
\put(210,37){$\mrm{q}$}
\put(210,57){$\mrm{\bar{q}}$}
\Photon(100,20)(150,40){-3}{4.5}
\Photon(100,80)(150,60){3}{4.5}
\put(85,77){$\gamma^{*}$}
\put(85,17){$\gamma$}
\end{picture}
\end{center}
\caption{The high momentum implies that the quarks move parallel to 
each other.}
\label{fig:quarks}
\end{figure}

The third limit is when one of the photons is real and the other one 
very much off shell, e.g.\ $k_{1}^{2}=0$ and $k_{2}^{2}<<0$. Which one 
is zero doesn't matter since the form-factor has to be symmetric in the 
two photons. In this case
% the pseudo-scalar created will have high momentum 
one can use that this is an exclusive process at high momentum transfer
 and the transition 
can be seen like in Fig.~\ref{fig:quarks} where the quark and anti-quark
that builds up the meson move parallel to each other.
%
%****************Figure of Parallel Quarks*****************************
%***********************************************************************
%
This simplifies the calculations and the answer from QCD is 
\begin{displaymath}
F(0,k^{2})=-\frac{C\ 8\pi^{2}\ F_{\pi}^{2}}{N_{c}\ k^{2}},\ \ \ \ \ 
k^{2}_{1}=0, k^{2}_{2}=k^{2}<<0.
\end{displaymath}
where $C$ is a constant that is $1$, $2$ or $3$ depending on further 
physical assumptions \cite{ref:Hayakawa}.

That $F(k^{2}_{1},k^{2}_{2})$ goes like $1/k^{2}$ in these two cases is
very reliable. The next terms are of the order $1/k^{4}$ so they are 
obviously much smaller. The coefficients in front are a bit more 
uncertain though, since corrections to them are of order $\log(k^{2})$ 
and that is not so small. This will be referred to in the next section.

The last constraint is not of the same form as the others. It is just 
that one believes that the quark and gluon effects which are not 
included in the pointlike treatment should somehow correspond to 
intermediate states with other mesons, like $\rho$. This idea comes 
among other things from the fact that the similar reaction
$e^{+}e^{-}\rightarrow\gamma^{*}\rightarrow\pi^{+}\pi^{-}$ is very 
well explained in a picture with an intermediate state like 
$e^{+}e^{-}\rightarrow\gamma^{*}\rightarrow\rho\rightarrow
\pi^{+}\pi^{-}$. This means that the shape of the form-factor should 
be such that it can be explained within the picture of an intermediate 
state vector meson. This is not absolutely certain since one 
doesn't really know the QCD-calculations, but there are some other
theoretical arguments that point in this direction as well
\cite{ref:Sakurai}. 

%***********************************************************************
\subsection{The Form-factors in this Article}   
\label{subsec-Form here}
%***********************************************************************

It proves very hard to find a form-factor that satisfies these four 
constraints. Actually none of the factors we study in this article
does that. We have chosen four different form-factors, each satisfying
some of the constraints, but not all. Nevertheless, it is still 
possible to see how much the choice of form-factor influences 
different physical quantities.

The first form-factor is just $F(k^{2}_{1},k^{2}_{2})=1$. That means no
form-factor, and the reaction is seen as pointlike. This form-factor 
obviously satisfies the first constraint but not the others. This is no
 realistic form but it can be studied to see how much the introduction
 of a form-factor means.

The next form-factor is constructed to reproduce the experimental data
that exist in this area. Experiments has been done with one photon 
real and the other off shell \cite{ref:exp}. All our form-factors will 
be compared to these data in the next section. This form-factor is
\begin{displaymath}
F(k^{2}_{1},k^{2}_{2})=\displaystyle{
\frac{m_{\rho}^{4}}{(m_{\rho}^{2}-k_{1}^{2})(m_{\rho}^{2}-k_{2}^{2})}},
\end{displaymath} 
where $m_{\rho}$ is the mass of the $\rho$-meson. It satisfies the first
and fourth requirement and with $F_{\pi}=92.4\,\mrm{MeV}$ and 
$m_{\rho}=770\,\mrm{MeV}$ it satisfies the third constraint within the 
uncertainty mentioned above.

The third form-factor we have made up ourselves to satisfy three of the
four constraints and it looks like
\begin{displaymath}
F(k^{2}_{1},k^{2}_{2})=\displaystyle{
\frac{m_{\rho}^{2}}{(m_{\rho}^{2}-k_{1}^{2}-k_{2}^{2})}}.
\end{displaymath}    
It satisfies constraint one, two and three with the same comment as
the last one, but not four, it can't be explained in terms of 
intermediate vector mesons.

The last one is also constructed to satisfy a specific requirement, 
but this time number two. It is
\begin{displaymath}
F(k^{2}_{1},k^{2}_{2})=\displaystyle{
\frac{m_{\rho}^{4}-\frac{4\pi^{2}\ F_{\pi}^{2}}{N_{c}}\ (k_{1}^{2}+
k_{2}^{2})}{(m_{\rho}^{2}-k_{1}^{2})(m_{\rho}^{2}-k_{2}^{2})}},
\end{displaymath}
and it satisfies constraint one, two and four.\\

To sum up, the form-factors we are looking at in this article are:
%
%************Array of form-factors**************************************
\\
\\
\begin{displaymath}
\begin{array}{ll}
\mathit{\mrm{Form\ Factor\ 1:}} & F(k^{2}_{1},k^{2}_{2})=1 \\ 
\\
\mathit{\mrm{Form\ Factor\ 2:}} & F(k^{2}_{1},k^{2}_{2})=\displaystyle{
\frac{m_{\rho}^{4}}{(m_{\rho}^{2}-k_{1}^{2})(m_{\rho}^{2}-k_{2}^{2})}}\\
\\
\mathit{\mrm{Form\ Factor\ 3:}} & F(k^{2}_{1},k^{2}_{2})=\displaystyle{
\frac{m_{\rho}^{2}}{(m_{\rho}^{2}-k_{1}^{2}-k_{2}^{2})}} \\
\\
\mathit{\mrm{Form\ Factor\ 4:}} & F(k^{2}_{1},k^{2}_{2})=\displaystyle{
\frac{m_{\rho}^{4}-\frac{4\pi^{2}\ F_{\pi}^{2}}{N_{c}}\ (k_{1}^{2}+
k_{2}^{2})}{(m_{\rho}^{2}-k_{1}^{2})(m_{\rho}^{2}-k_{2}^{2})}} \\
\end{array}
\end{displaymath}
%***********************************************************************
%
\clearpage

%***********************************************************************
\subsection{Comparison with Experiment}   
\label{subsec-Exp}
%***********************************************************************

\begin{figure}[t]
\begin{center}
%\rotatebox{270}{\mbox{\epsfig{file=formexp.eps,height=15cm}}}
%\epsfig{file=formexp.eps,height=15cm, angle=270}
%\epsfig{file=formexp.eps, height=0.95\textwidth, angle=270}
\epsfig{file=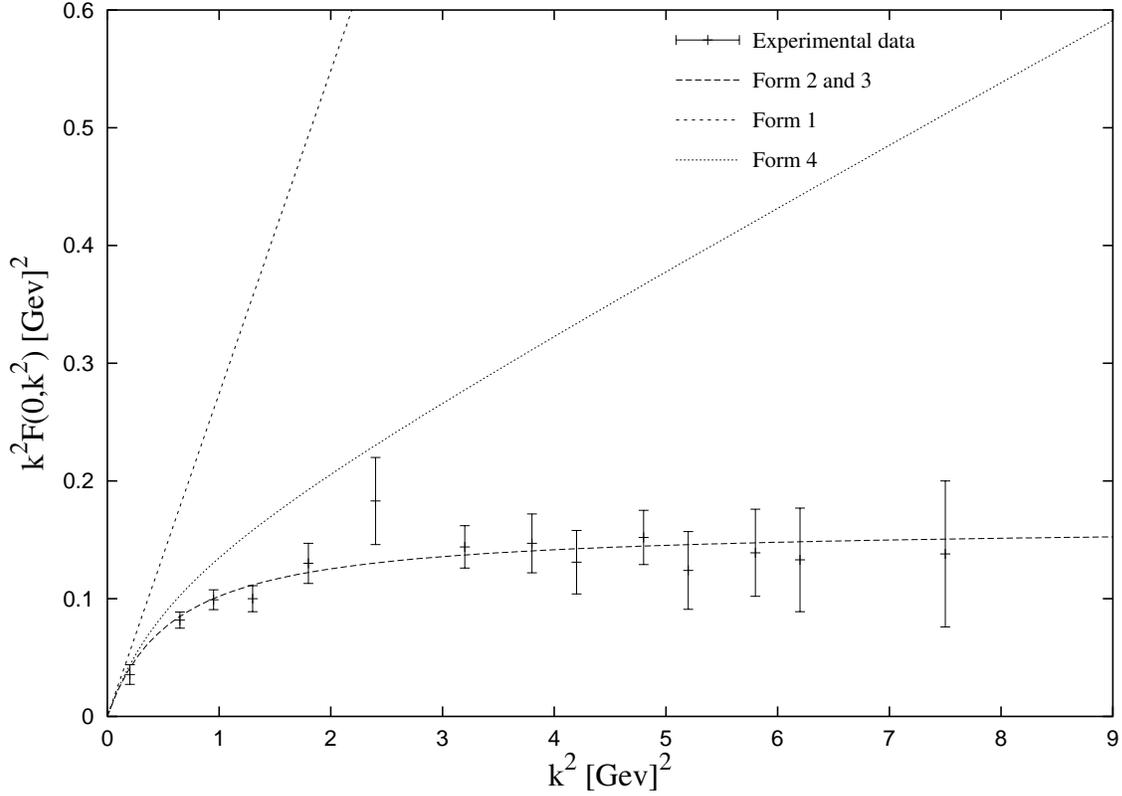,angle=270,width=0.95\textwidth}
\end{center}
\caption{Comparison between our four form-factors and data from the CLEO
 collaboration.}
\label{fig:formexp} 
\end{figure}

Experiments with form-factors has been done by the CLEO 
collaboration \cite{ref:exp}. They have studied the pion form-factor 
in the case that one of the photons is real and the other one is 
virtual. In this case our Form 2 and Form 3 have the same shape. In 
Fig.~\ref{fig:formexp} the form-factors presented in the last section
are compared with the data from the experiment. Their normalization
of the form-factors is not the same as ours. Therefore our form-factors
from the last section are multiplied with a factor
\begin{displaymath}
\sqrt{\frac{64\pi\ \Gamma(\pi^{0}\rightarrow\gamma\gamma)}
{(4\pi\alpha)^{2}\ m_{\pi^{0}}^{3}}}.
\end{displaymath}

It is easy to see that Form 1 and Form 4 do not satisfy the data as 
well as the other two. But still we will study the results from them 
to get a feeling for how much influence an incorrect form 
factor will have on various results.  
%
%***********Figure of Experimental Data*********************************
%***********************************************************************
%
\clearpage

%***********************************************************************

% SECTION (Decay Rates)

\section{Decay Rates}
\label{sec-Decay}

%***********************************************************************

The decay rate of a particle is related to its lifetime. One version
of Heisenbergs uncertainty relation can be written
\begin{equation}
\Delta E\ \Delta t\simeq \hbar,
\label{eq:rel} 
\end{equation}
where $\Delta t$ can be seen as the lifetime of a particle. 
If a particle has a finite lifetime, Eq.~(\ref{eq:rel}) states that
it must have an uncertainty in energy. This uncertainty is usually
called the decay width, or the decay rate, of the particle.
  
In this section the decay rates of three processes, which include the  
meson-photon-photon vertex $\eta\rightarrow\gamma\gamma$, 
is calculated. The reason for choosing $\eta$ is that an upcoming 
experiment, WASA at CELSIUS, Uppsala, is going to produce up to 
$10^{10}$ $\eta$-particles. 
Therefore, it is interesting to conclude whether they will 
be able to see the difference between the form-factors. We will present
results for our four form-factors to see whether they can be 
distinguished in the experiment. But first we study $\pi^{0}$ decay to 
check our approximations.    

%***********************************************************************
\subsection{$\pi^{0}\rightarrow\gamma\gamma$}   
\label{subsec-vertex}
%***********************************************************************

First we will study this process where $\pi^{0}$ decays. There is no 
dependence on form-factors since both photons are real, but 
it is still interesting to study it to check the 
different approximations we have made. The answer can be compared with
the experimental lifetime of the pion and thus we have a measure
of how well our calculations work.

%***********************************************************************
\subsubsection{The Amplitude}
\label{subsubsec-amp}
%***********************************************************************

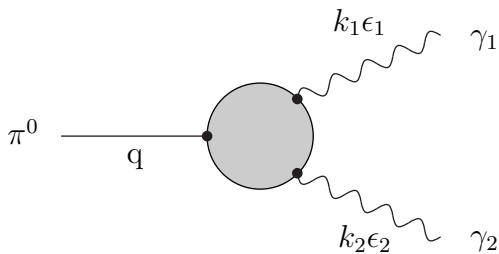
\begin{figure}[b]
\begin{center}
\begin{picture}(300,100)
\GCirc(150,50){20}{0.8}
\Vertex(130,50){2}
\Vertex(164,64){2}
\Vertex(164,36){2}
\Photon(164,64)(218,88){3}{4.5}
\Photon(164,36)(218,12){-3}{4.5}
\Line(75,50)(130,50)
\put(100,40){q}
\put(178,90){$k_{1}\epsilon_{1}$}
\put(180,8){$k_{2}\epsilon_{2}$}
\put(55,47){$\pi^{0}$}
\put(230,85){$\gamma_{1}$}
\put(230,9){$\gamma_{2}$}
\end{picture}
\end{center}
\caption{The decay of $\pi^{0}$ into two photons.}
\label{fig:pion}
\end{figure}

The process, as can be seen in Fig.~\ref{fig:pion}, includes only the 
triangle vertex itself and 
no propagators. The triangle vertex is described by the Lagrangian
\begin{equation}
\mathcal{L}=-\frac{e^{2}}{4\sqrt{2}\pi^{2}F_{\pi}}\ \partial_{\mu}{A_{\nu}(x)}\ A_{\alpha}(x)\ \partial_{\beta}{\phi(x)}\ \varepsilon^{\mu\nu\alpha\beta}.
\label{eq:Lagrange} 				
\end{equation}
%
%
%****************Figure of Decay of Pion********************************
%***********************************************************************
%
Here $\phi(x)$ is the proper mixture of the meson fields,
\begin{equation}
\phi(x)=\frac{1}{\sqrt{2}}\pi^{0}(x)+\frac{1}{\sqrt{6}}\eta(x)+
\frac{2}{\sqrt{3}}\eta'(x),
\end{equation}
where e.g. the pion field is written as 
\begin{equation}
\pi^{0}(x)=\int N_{\pi^{0}}(q)\ (e^{-iqx}a_{q}+e^{iqx}a_{q}^{\dagger})
\ d^{3}q.
\end{equation}
The operators $a_{q}$ and $a_{q}^{\dagger}$ annihilate and create, 
respectively, a pion with momentum $q$ and $N_{\pi^{0}}(q)$ is a 
numerical factor which is taken care of by the Feynman rules.
The other two meson fields
 can be written in the same way. 

The photon field needs
an extra index since it describes a vector particle and not a scalar. 
This is introduced as a polarization vector $\epsilon$, which has four 
components. The photon field is then
\begin{equation}
A_{\mu}(x)=\int N_{A}(q)\ (\epsilon_{\mu}e^{-ikx}a_{k}+
\epsilon_{\mu}^{\ast}
e^{ikx}a_{k}^{\dagger})\ d^{3}q.
\end{equation}
In the process we are looking at one pion is annihilated and two 
photons are created. The corresponding terms are picked out from the 
field operators and used in 
the Lagrangian to get the Feynman amplitude for the vertex. This gives
\begin{equation}
A(\pi^{0}\rightarrow\gamma\gamma)=\frac{e^{2}}{8\pi^{2}F_{\pi}}
\ \varepsilon^{\mu\nu\alpha\beta}\ q_{\beta}\ \epsilon_{1\nu}^{\ast}
k_{1\mu}\epsilon_{2\alpha}^{\ast}.
\end{equation}
But this is not the final answer. According to the Feynman rules
an extra $i$ should be added and furthermore, the photons are 
identical particles. This means that another term have 
to be added to the amplitude, in which the substitution 
$\gamma_{1}\leftrightarrow\gamma_{2}$ is made. This gives the result
\begin{equation}
A(\pi^{0}\rightarrow\gamma\gamma)=\frac{ie^{2}}{8\pi^{2}F_{\pi}}
\ \varepsilon^{\mu\nu\alpha\beta}\ q_{\beta}\ [\epsilon_{1\nu}^{\ast}
k_{1\mu}\epsilon_{2\alpha}^{\ast}+\epsilon_{2\nu}^{\ast}k_{2\mu}
\epsilon_{1\alpha}^{\ast}].
\end{equation}
\\

%***********************************************************************
\subsubsection{The Matrix Element Squared}
\label{subsubsec-matrix}
%***********************************************************************

But to calculate the decay rate we need the amplitude squared, usually 
called the matrix element squared,
\begin{equation}
|M|^{2}=|A|^{2}=A\ A^{\ast}.
\end{equation}
When this is calculated, one has to sum over all polarization states. 
This is taken care of by using the rule
\begin{equation}
\sum \epsilon_{1\nu}\epsilon_{1\tau}^{\ast}=-g_{\nu\tau},
\end{equation}
where the sum goes over the two polarization states of the photons and 
$g_{\nu\tau}$ is the metric tensor. Rules for the Levi-Civita tensor
such as
\begin{equation}
\varepsilon^{\mu\nu\alpha\beta}\ \varepsilon^{\sigma\ \ \rho}
_{\ \nu\alpha\ }=-2(g^{\mu\sigma}g^{\beta\rho}-g^{\mu\rho}
g^{\sigma\beta})
\end{equation}
also have to be used. 

The matrix element includes terms of scalar products of all the momenta 
in the process. These scalar products have to be calculated and put in. 
The result is
%
%*******************Array of Matrix Squared*****************************
\begin{equation}
|M|^{2}(\pi^{0}\rightarrow\gamma\gamma)=\displaystyle{\frac{e^{4}}
{2^{6}\pi^{4}F^{2}_{\pi}}\ m^{4}_{\pi^{0}}}.
\label{eq:matsq} 
\end{equation}
%***********************************************************************
%
\clearpage

%***********************************************************************
\subsubsection{The decay rate}
\label{subsubsec-decay}
%***********************************************************************

Now the matrix element squared should be connected to the decay rate. 
This is done using the general formula \cite{ref:Journal}
\begin{equation}
d\Gamma=\frac{(2\pi)^{4}}{2\mrm{M}}\ |M|^{2}\ d\Phi_{n},
\label{eq:Gamma} 
\end{equation}
where $\mrm{M}$ is the mass of the decaying particle, $n$ the 
number of particles in the final state and
\begin{equation}
d\Phi_{n}=\delta^{4}(\mathbf{P}-\sum_{i=1}^{n} \mathbf{p_{i}})\ 
\prod_{i=1}^{n} \frac{d^{3}\vec{p}_{i}}{(2\pi)^{3}\ 2E_{i}}
\end{equation}
In this case there are two degrees of freedom and the formula just gives
\begin{equation}
d\Gamma=\frac{|M|^{2}\ |\mathbf{k}|}{32\pi^{2}\mrm{M^{2}}}\ d\Omega
\end{equation}
where $|\mathbf{k}|$ is the magnitude of either of the photons four
momenta, equal for both. 

In the case we are looking at $|\mathbf{k}|=m_{\pi}/2$, and since the 
two photons are identical and $\pi^{0}$ is a pseudo-scalar, 
$d\Omega=2\pi$. The final result is then
%
%*****************Array of Decay Rates**********************************
\begin{displaymath}
\Gamma(\pi^{0}\rightarrow\gamma\gamma)=\displaystyle{\frac{\alpha^{2}\ 
m_{\pi^{0}}^{3}}{2^{6}\pi^{3}F_{\pi}^{2}}}
\end{displaymath}
%***********************************************************************
%
with $\alpha=e^{2}/4\pi$.

%***********************************************************************
\subsubsection{The Result}
\label{subsubsec-res}
%***********************************************************************

Putting in the numbers
%
%*************Array of Numbers******************************************
\begin{equation}
\begin{array}{lll}
F_{\pi}=92.4\ MeV & \alpha=1/137 & m_{\pi^{0}}=134.98\ MeV \\
%m_{\pi^{0}}=134.98\ MeV & m_{\eta}=547.5\ MeV & m_{\eta'}=957.8\ MeV,
\end{array}
\end{equation}
%***********************************************************************
%
gives the numerical result
%
%**************Array of Decay Rate Results******************************
\begin{equation}
%\begin{array}{ll}
\Gamma(\pi^{0}\rightarrow\gamma\gamma)= 7.734\times10^{-6}\ MeV  \\ \\
%\Gamma(\eta\rightarrow\gamma\gamma)= & 1.720\times10^{-4} \\ \\
%\Gamma(\eta'\rightarrow\gamma\gamma)= & 3.684\times10^{-3}.
%\end{array}
\end{equation}
%***********************************************************************
%

Knowing how often the pion decays in this channel, it is possible 
to calculate its lifetime, $\tau$.
%
%************Array of Lifetimes*****************************************
\begin{displaymath}
%\begin{array}{llll}
\tau(\pi^{0})= \displaystyle{\frac{0.988}{\Gamma}}=127800\ 
MeV^{-1}=8.41\times10^{-17}\ s \\ \\
%\tau(\eta)= & \displaystyle{\frac{0.393}{\Gamma}}= & 2284\ 
%MeV^{-1}= & 1.50\times10^{-18}\ s \\ \\
%\tau(\eta')= & \displaystyle{\frac{0.021}{\Gamma}}= & 5.700\ 
%MeV^{-1}= & 3.75\times10^{-21}\ s
%\end{array}
\end{displaymath}
%***********************************************************************
%
Experiments give
\begin{displaymath}
\begin{array}{ll}
\tau(\pi^{0})= & 8.4\times10^{-17}\ s
\end{array}
\end{displaymath}
so our approximations seem valid.

This is expected to be correct for the pion but wouldn't have been
as exact for $\eta$ and $\eta'$ since they 
are more complicated. First of all the s quark is much heavier than the 
u and d quark, which make our approximations more severe for $\eta$
and $\eta'$. Furthermore their mass is affected by gluon effects which
we haven't included.

%***********************************************************************
\subsection{$\eta\rightarrow\gamma\ e^{+} e^{-}$ and 
$\eta\rightarrow\gamma\ \mu^{+} \mu^{-}$}   
\label{subsec-gammaee}
%***********************************************************************

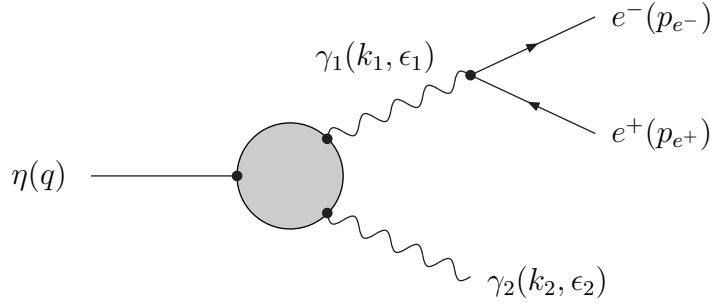
\begin{figure}[t]
\begin{center}
\begin{picture}(300,150)
\GCirc(130,50){20}{0.8}
\Vertex(110,50){2}
\Vertex(144,64){2}
\Vertex(144,36){2}
\Vertex(198,88){2}
\Photon(144,64)(198,88){3}{4.5}
\Photon(144,36)(198,12){-3}{4.5}
\Line(55,50)(110,50)
\ArrowLine(198,88)(245,110)
\ArrowLine(245,66)(198,88)
\put(252,63){$e^{+}(p_{e^{+}})$}
\put(252,107){$e^{-}(p_{e^{-}})$}
%\put(80,40){q}
\put(140,92){$\gamma_{1}(k_{1},\epsilon_{1})$}
\put(205,7){$\gamma_{2}(k_{2},\epsilon_{2})$}
\put(25,47){$\eta(q)$}
%\put(210,85){$\gamma_{1}$}
%\put(210,9){$\gamma_{2}$}
\end{picture}
\end{center}
\caption{The decay of $\eta$ into photon, electron and positron.}
\label{fig:gammaee}
\end{figure}

The process with the electron and positron is described in 
Fig.~\ref{fig:gammaee} 
and it is the first 
case in which we can study the effects of the form 
factors. Here a new vertex is added which is described by the Lagrangian
\begin{equation}
\mathcal{L}=-e\ (\bar{\mrm{e}}\gamma^{\mu}\mrm{e})\ A_{\mu},
\end{equation}
%
%
%**************Figure of Gamma e e**************************************
%***********************************************************************
%
where $\bar{\mrm{e}}$ and $\mrm{e}$ are the positron and electron fields
 and $\gamma$ stands for the $\gamma$-matrices. There is also a photon
propagator, which according to the Feynman rules is described by 
$-i/k_{1}^{2}$ in this case. These two factors get multiplied to 
Eq.~(\ref{eq:Lagrange}) which then leads to a new amplitude
\begin{equation}
A=-\frac{e^{3}}{8\sqrt{3}\pi^{2}\ F_{\pi}}\ \varepsilon^
{\mu\nu\alpha\beta}\
\frac{q_{\beta}}{k_{1}^{2}}\ [(\bar{\mrm{e}}\gamma_{\alpha}\mrm{e})
\epsilon_{1\nu}^{\ast}k_{1\mu}+(\bar{\mrm{e}}\gamma_{\nu}\mrm{e})
\epsilon_{1\alpha}^{\ast}k_{2\mu}].
\end{equation}
This is then squared, which e.g.\ includes working out traces of 
the $\gamma$-matrices. The result was checked with the 
program FORM. The 
matrix element squared is then used in the kinematics formula for 
three body
decay, coming from Eq.~(\ref{eq:Gamma}),
\begin{equation}
d\Gamma=\frac{|M|^{2}}{(2\pi)^{3}\ 32\mrm{M}^{3}_{PS}}\ 
dm^{2}_{e^{+}e^{-}}\ dm^{2}_{e^{-}\gamma},
\end{equation} 
where $m^{2}_{e^{+}e^{-}}=(p_{e^{+}}+p_{e^{-}})^{2}$ and 
$m^{2}_{e^{-}\gamma}=(p_{e^{-}}+k_{2})^{2}$ are the two degrees of 
freedom we have chosen. These are then integrated over, 
$dm^{2}_{e^{-}\gamma}$ analytically and $dm^{2}_{e^{+}e^{-}}$ using the
Gaussian quadrature program DGAUSS. In the last integration different
cuts on the lower limit are made to see if this gives more difference 
between the decay 
rates for different form-factors. The decay rate is then divided by 
\begin{displaymath}
\Gamma^{\ast}(\eta\rightarrow\gamma\gamma)=1.720\times10^{-4}\ MeV,
\end{displaymath}
where $\ast$ means that it is our \emph{theoretical} value of this 
decay rate.
We use this one because of the arguments before that our approximations
are not quite correct in the $\eta$ and $\eta'$ case. In this way, the 
same error should exist in both numerator and denominator so the 
ratio will be a more accurate number.  
The result is shown in Table~\ref{tab:ee1}, with numerical errors 
smaller than the last digit in the quoted values. The last line is the
value without cut for $\eta\rightarrow\gamma\ \mu^{+} \mu^{-}$. We have 
used $M_{\eta}=547.5\ MeV$.

%***********************************************************************
\subsubsection{Experimental Uncertainties}
%***********************************************************************

It is necessary here to get a feeling for how large the uncertainty
will be in the forthcoming experiment. A first estimate of the error is
$\sqrt{N}$, where $N$ is the number of decaying particles.
$N$ is given by
%
%************Table of Different Form I**********************************
\begin{table}[t]
\begin{center}
\begin{tabular}{||c|r|r|r||} \hline
cut\ ($MeV^{2}$) & Form 1 & Form 2,3 & Form 4 \\ \hline
0 & $1.620\times10^{-2}$&$1.666\times10^{-2}$&$1.657\times10^{-2}$\\
2500&$4.614\times10^{-3}$&$5.071\times10^{-3}$&$4.980\times10^{-3}$\\
10000&$2.580\times10^{-3}$&$3.000\times10^{-3}$&$2.916\times10^{-3}$\\
22500&$1.507\times10^{-3}$&$1.869\times10^{-3}$&$1.796\times10^{-3}$\\
40000&$8.601\times10^{-4}$&$1.150\times10^{-3}$&$1.091\times10^{-3}$\\
62500&$4.614\times10^{-4}$&$6.748\times10^{-4}$&$6.308\times10^{-4}$\\
90000&$2.237\times10^{-4}$&$3.636\times10^{-4}$&$3.342\times10^{-4}$\\
122500&$9.258\times10^{-5}$&$1.705\times10^{-4}$&$1.539\times10^{-4}$\\
160000&$2.958\times10^{-5}$&$6.322\times10^{-5}$&$5.585\times10^{-5}$\\
202500&$5.825\times10^{-6}$&$1.488\times10^{-5}$&$1.284\times10^{-5}$\\
250000&$3.394\times10^{-7}$&$1.077\times10^{-6}$&$9.050\times10^{-7}$\\
\hline
$\gamma \mu^{+} \mu^{-}$&$5.506\times10^{-4}$&$7.744\times10^{-4}$&
$7.284\times10^{-4}$\\
\hline
\end{tabular}
\caption{Decay rate for $\eta\rightarrow\gamma\ e^{+}e^{-}$, normalized
to $\Gamma (\eta\rightarrow\gamma\gamma)$.}
\label{tab:ee1}
\end{center}  
\end{table}
%***********************************************************************
%
\begin{equation}
N=N_{0}\ \frac{\Gamma}{\Gamma_{\gamma\gamma}}\ 
\frac{\Gamma_{\gamma\gamma}}{\Gamma_{0}}=N_{0}\ 
\frac{\Gamma}{\Gamma_{\gamma\gamma}}\ 
BR(\eta\rightarrow\gamma\gamma),
\end{equation}  
where $\Gamma_{0}$ is the total decay rate and 
$\Gamma/\Gamma_{\gamma\gamma}$ the number given in the table. From this 
one can calculate the uncertainty on the normalized decay rate,
\begin{equation}
\frac{\Gamma}{\Gamma_{\gamma\gamma}}\pm\sqrt{\frac{
\Gamma/\Gamma_{\gamma\gamma}}{BR(\eta\rightarrow\gamma\gamma)\
N_{0}}}
\label{eq:error} 
\end{equation}
Without any cut and with $N_{0}=10^{10}$ and 
\begin{displaymath}
BR(\eta\rightarrow\gamma\gamma)=0.39,
\end{displaymath}
this gives an uncertainty $\sigma\approx 10^{-6}$. 
So it should be possible to discern the different form-factors here.

But then a cut is introduced to see whether the difference between the 
form-factors will increase.
For the cut on $160000\ MeV^{2}$, the difference between the 
form-factors is quite sizeable. From Eq.~(\ref{eq:error}) one sees that
$\sigma\approx 10^{-7}$ in this case, which means that it should still 
be possible to distinguish between the three different form-factors 
with a high cut in the experiment. The reaction 
$\eta\rightarrow\gamma\ \mu^{+} \mu^{-}$ gives the same result. 

It remains to check whether it is possible to see the 
difference between Form 2 and 3 in some other decay.

%***********************************************************************
\subsection{$\eta\rightarrow\mu^{+}\mu^{-}\ e^{+} e^{-}$}   
\label{subsec-uuee}
%***********************************************************************
 
\begin{figure}[t]
\begin{center}
\begin{picture}(300,150)
\GCirc(130,70){20}{0.8}
\Vertex(110,70){2}
\Vertex(144,84){2}
\Vertex(144,56){2}
\Vertex(198,108){2}
\Vertex(198,32){2}
\Photon(144,84)(198,108){3}{4.5}
\Photon(144,56)(198,32){-3}{4.5}
\Line(55,70)(110,70)
\ArrowLine(198,108)(245,130)
\ArrowLine(245,86)(198,108)
\ArrowLine(198,32)(245,54)
\ArrowLine(245,10)(198,32)
\put(252,83){$e^{+}(p_{e^{+}})$}
\put(252,127){$e^{-}(p_{e^{-}})$}
\put(252,7){$\mu^{+}(p_{\mu^{+}})$}
\put(252,51){$\mu^{-}(p_{\mu^{-}})$}
\put(140,112){$\gamma_{1}(k_{1},\epsilon_{1})$}
\put(140,22){$\gamma_{2}(k_{2},\epsilon_{2})$}
\put(25,67){$\eta(q)$}
\end{picture}
\end{center}
\caption{The decay of $\eta$ into muon, anti-muon, electron and 
positron.}
\label{fig:uuee}
\end{figure}
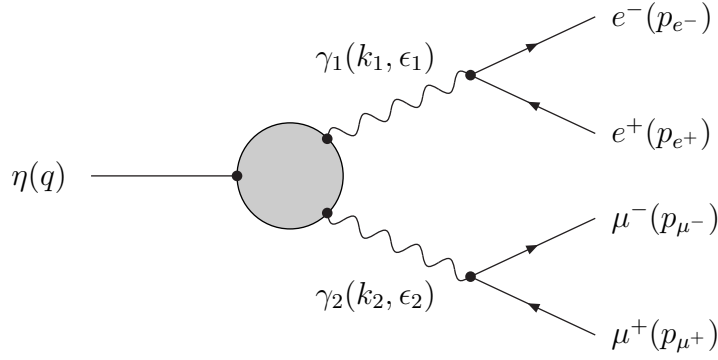

As one can see in Fig.~\ref{fig:uuee}, one new vertex and one new 
propagator is added to the last reaction. This leads to an amplitude
\begin{equation}
A=-\frac{ie^{4}}{8\sqrt{3}\pi^{2}\ F_{\pi}}\ \varepsilon^
{\mu\nu\alpha\beta}\
\frac{q_{\beta}}{k_{1}^{2}k_{2}^{2}}\ [(\bar{\mrm{e}}\gamma_{\alpha}
\mrm{e})k_{1\mu}(\bar{\mrm{\mu}}\gamma_{\nu}
\mrm{\mu})+(\bar{\mrm{e}}\gamma_{\nu}\mrm{e})k_{2\mu}(\bar{\mrm{\mu}}
\gamma_{\alpha}\mrm{\mu})].
\label{eq:ampeeuu}
\end{equation}
Calculating the corresponding matrix element squared includes 
working out the Dirac traces and once again the result is checked with 
FORM.
 The matrix element is then inserted in
the formula for the four body decay rate, which has five degrees of 
freedom,
%**************Figure of u u e e****************************************
%***********************************************************************
%
%
\begin{equation}
d\Gamma=\displaystyle{\frac{|M|^{2}|\vec{k_{1}}||\vec{p_{e}}^{\ast}||
\vec{p_{\mu}}^{\ast\ast}|}{(2\pi)^{6}64\mrm{M^{2}}m_{1}m_{2}}}\ 
d(\cos{\theta_{1}}^{\ast})\ d(\cos{\theta_{2}}^{\ast\ast})\ 
d\phi_{2}^{\ast\ast}\ dm_{1}^{2}\ dm_{2}^{2},
\label{eq:4body}
\end{equation}
where $m_{1}=\sqrt{(p_{e^{+}}+p_{e^{-}})^{2}},\ 
m_{2}=\sqrt{(p_{\mu^{+}}+p_{\mu^{-}})^{2}}$ and $\ast(\ast\ast)$ 
indicates that the quantity should be calculated in the rest frame of
$\gamma_{1}(\gamma_{2})$. These five degrees of freedom are again
integrated over, this time the angle integrations are done analytically 
and the other two with DGAUSS.
This leads to different results for the form-factors 
according to Table~\ref{tab:eeuu1}. Once again cuts are made, this time 
we impose a lower limit on $m_{1}^{2}$, to see if this enhances the 
difference between the factors.
% 
%*************Table of different form II********************************
\begin{table}[h]
\begin{center}
\begin{tabular}{||c|r|r|r|r||} \hline
cut\ ($MeV^{2}$) & Form 1 & Form 2 & Form 3 & Form 4 \\ \hline
0 & $3.99\times10^{-6}$&$5.61\times10^{-6}$&$5.62\times10^{-6}$
&$5.28\times10^{-6}$ \\
625&$1.39\times10^{-6}$&$1.96\times10^{-6}$&$1.97\times10^{-6}$
&$1.84\times10^{-6}$\\
2500&$8.26\times10^{-7}$&$1.16\times10^{-6}$&$1.17\times10^{-6}$
&$1.09\times10^{-6}$\\
5625&$5.21\times10^{-7}$&$7.37\times10^{-7}$&$7.44\times10^{-7}$
&$6.92\times10^{-7}$\\
10000&$3.29\times10^{-7}$&$4.68\times10^{-7}$&$4.74\times10^{-7}$
&$4.39\times10^{-7}$\\
15625&$2.03\times10^{-7}$&$2.91\times10^{-7}$&$2.96\times10^{-7}$
&$2.73\times10^{-7}$\\
22500&$1.20\times10^{-7}$&$1.74\times10^{-7}$&$1.77\times10^{-7}$
&$1.63\times10^{-7}$\\
30625&$6.64\times10^{-8}$&$9.79\times10^{-8}$&$9.99\times10^{-8}$
&$9.12\times10^{-8}$\\
40000&$3.36\times10^{-8}$&$5.04\times10^{-8}$&$5.16\times10^{-8}$
&$4.68\times10^{-8}$\\
50625&$1.49\times10^{-8}$&$2.28\times10^{-8}$&$2.34\times10^{-8}$
&$2.11\times10^{-8}$\\
62500&$5.36\times10^{-9}$&$8.46\times10^{-9}$&$8.70\times10^{-9}$
&$7.78\times10^{-9}$\\
75625&$1.37\times10^{-9}$&$2.22\times10^{-9}$&$2.29\times10^{-9}$
&$2.03\times10^{-9}$\\
90000&$1.68\times10^{-10}$&$2.82\times10^{-10}$&$2.92\times10^{-10}$
&$2.57\times10^{-10}$\\
\hline
\end{tabular}
\caption{Decay rate for $\eta\rightarrow\mu^{+}\mu^{-}\ e^{+} e^{-}$,
normalized to $\Gamma (\eta\rightarrow\gamma\gamma)$.}
\label{tab:eeuu1}
\end{center}
\end{table}
%***********************************************************************
%
For the lowest cut, $625\ MeV^{2}$, the uncertainty calculated from 
Eq.~(\ref{eq:error})
is $\sigma\approx 10^{-8}$. Then it is possible to distinguish 
Form~1 and Form~4 from the others. For a higher cut, like 
$30625\ MeV^{2}$, 
the error should be $\sigma\approx 5\times10^{-9}$. It still is not
possible to see the difference between Form 2 and Form 3.\\

%***********************************************************************
\subsection{$\eta\rightarrow e^{+} e^{-}\ e^{+} e^{-}$}
%***********************************************************************
%
%**************Figure of e e e e****************************************
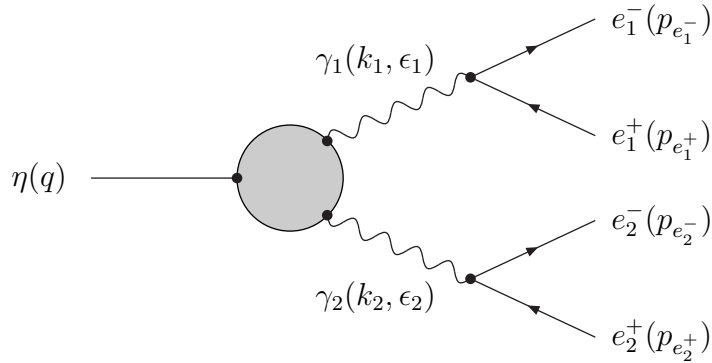
\begin{figure}[b]
\begin{center}
\begin{picture}(300,150)
\GCirc(130,70){20}{0.8}
\Vertex(110,70){2}
\Vertex(144,84){2}
\Vertex(144,56){2}
\Vertex(198,108){2}
\Vertex(198,32){2}
\Photon(144,84)(198,108){3}{4.5}
\Photon(144,56)(198,32){-3}{4.5}
\Line(55,70)(110,70)
\ArrowLine(198,108)(245,130)
\ArrowLine(245,86)(198,108)
\ArrowLine(198,32)(245,54)
\ArrowLine(245,10)(198,32)
\put(252,83){$e^{+}_{1}(p_{e^{+}_{1}})$}
\put(252,127){$e^{-}_{1}(p_{e^{-}_{1}})$}
\put(252,7){$e^{+}_{2}(p_{e^{+}_{2}})$}
\put(252,51){$e^{-}_{2}(p_{e^{-}_{2}})$}
\put(140,112){$\gamma_{1}(k_{1},\epsilon_{1})$}
\put(140,22){$\gamma_{2}(k_{2},\epsilon_{2})$}
\put(25,67){$\eta(q)$}
\end{picture}
\end{center}
\caption{The decay of $\eta$ into two electrons and two positrons.}
\label{fig:eeee1}
\end{figure}
\begin{figure}[t]
\begin{center}
\begin{picture}(300,150)
\GCirc(130,70){20}{0.8}
\Vertex(110,70){2}
\Vertex(144,84){2}
\Vertex(144,56){2}
\Vertex(198,108){2}
\Vertex(198,32){2}
\Photon(144,84)(198,108){3}{4.5}
\Photon(144,56)(198,32){-3}{4.5}
\Line(55,70)(110,70)
\ArrowLine(198,108)(245,130)
\ArrowLine(245,86)(198,108)
\ArrowLine(198,32)(245,54)
\ArrowLine(245,10)(198,32)
\put(252,83){$e^{+}_{1}(p_{e^{+}_{1}})$}
\put(252,127){$e^{-}_{2}(p_{e^{-}_{2}})$}
\put(252,7){$e^{+}_{2}(p_{e^{+}_{2}})$}
\put(252,51){$e^{-}_{1}(p_{e^{-}_{1}})$}
\put(140,112){$\gamma_{3}(k_{3},\epsilon_{3})$}
\put(140,22){$\gamma_{4}(k_{4},\epsilon_{4})$}
\put(25,67){$\eta(q)$}
\end{picture}
\end{center}
\caption{The other electron configuration.}
\label{fig:eeee2}
\end{figure}
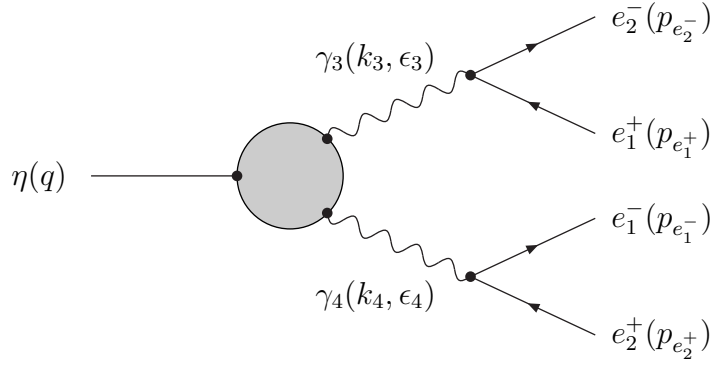

 
%***********************************************************************
%
The last decay we will study is the one depicted in 
Fig.~\ref{fig:eeee1}.
Kinematically, it is the same as the last one but the amplitude
gets more complicated. A start is to take the amplitude from the last
case, Eq.~(\ref{eq:ampeeuu})  and change $\mu$ to $e_{2}$. This gives
\begin{equation} 
A=-\frac{ie^{4}}{8\sqrt{3}\pi^{2}\ F_{\pi}}\ \varepsilon^
{\mu\nu\alpha\beta}\
\frac{q_{\beta}}{k_{1}^{2}k_{2}^{2}}\ [(\bar{\mrm{e}}_{1}\gamma_{\alpha}
\mrm{e_{1}})k_{1\mu}(\bar{\mrm{e}}_{2}\gamma_{\nu}
\mrm{e_{2}})+(\bar{\mrm{e}}_{1}\gamma_{\nu}\mrm{e_{1}})k_{2\mu}
(\bar{\mrm{e}}_{2}\gamma_{\alpha}\mrm{e_{2}})].
\label{eq:ampeeee}
\end{equation}
But this isn't enough, because here we once again have identical 
particles. Two pairs of two identical particles give four diagrams, 
which can be divided into pairs which give equal results.
The diagram in Fig.~\ref{fig:eeee1} represents one of the pairs and 
Fig.~\ref{fig:eeee2} represents the other pair.
%
%**************Figure of e e e e****************************************
%***********************************************************************
%
Both of these diagrams have to be taken into account and according to 
the Pauli principle they come in with different signs. Furthermore, 
this process is second order and this implies, according to the Feynman 
rules, an extra factor of $1/2$. 
Finally, the two pairs don't have
the same propagators. In the first case they are 
\begin{displaymath}
\begin{array}{l}
k_{1}=p_{e^{+}_{1}}+p_{e^{-}_{1}}\\
k_{2}=p_{e^{+}_{2}}+p_{e^{-}_{2}}
\end{array}
\end{displaymath}
but in the second case the propagators are
\begin{displaymath}
\begin{array}{l}
k_{3}=p_{e^{+}_{1}}+p_{e^{-}_{2}}\\
k_{4}=p_{e^{+}_{2}}+p_{e^{-}_{1}}.
\end{array}
\end{displaymath}
This gives an amplitude
\begin{equation}
A(\eta\rightarrow e^{+}e^{-}\ e^{+}e^{-})=-\frac{ie^{4}}{8\sqrt{3}
\pi^{2}\ F_{\pi}}\ \varepsilon^{\mu\nu\alpha\beta}
\ q_{\beta}\times[A_{1}-A_{2}]
\end{equation}
with
\begin{equation}
A_{1}=\frac{(\bar{\mrm{e}}_{1}\gamma_{\nu}
\mrm{e_{1}})k_{1\mu}(\bar{\mrm{e}}_{2}\gamma_{\alpha}
\mrm{e_{2}})+(\bar{\mrm{e}}_{1}\gamma_{\alpha}\mrm{e_{1}})k_{2\mu}
(\bar{\mrm{e}}_{2}\gamma_{\nu}\mrm{e_{2}})}{k_{1}^{2}k_{2}^{2}}
\end{equation}
and
\begin{equation} 
A_{2}=\frac{(\bar{\mrm{e}}_{1}\gamma_{\alpha}
\mrm{e_{2}})k_{3\mu}(\bar{\mrm{e}}_{2}\gamma_{\nu}
\mrm{e_{1}})+(\bar{\mrm{e}}_{1}\gamma_{\nu}\mrm{e_{2}})k_{4\mu}
(\bar{\mrm{e}}_{2}\gamma_{\alpha}\mrm{e_{1}})}{k_{3}^{2}k_{4}^{2}}.  
\end{equation}
This amplitude is then squared, using FORM, and put into 
Eq.~(\ref{eq:4body}), where
an extra factor $1/4$ is added since there are identical particles in
the final state, to get 
the decay rate. The five degrees of freedom are integrated over using
the Monte Carlo integration program VEGAS. Cuts are made on the low 
limit of $k_{1}^{2}$, $k_{2}^{2}$, $k_{3}^{2}$ and $k_{4}^{2}$, and 
the result is presented in Table~\ref{tab:eeee}. The errors in 
parantheses are the errors quoted from VEGAS. The errors on the numbers 
without parantheses is maximum one in the last digit.
%
%***************Table of Different Form III*****************************
\begin{table}[h]
\begin{center}
\begin{tabular}{||c|r|r|r|r||} \hline
cut\ ($MeV^{2}$) & Form 1 & Form 2 & Form 3 & Form 4 \\ \hline
0 & $6.40(2)\times10^{-5}$&$6.73(2)\times10^{-5}$&$6.71(2)\times10^{-5}$
&$6.64(2)\times10^{-5}$ \\
625&$9.93\times10^{-6}$&$1.10\times10^{-5}$&$1.10\times10^{-5}$
&$1.08\times10^{-5}$\\
2500&$3.83\times10^{-6}$&$4.40\times10^{-6}$&$4.41\times10^{-6}$
&$4.29\times10^{-6}$\\
5625&$1.57\times10^{-6}$&$1.86\times10^{-6}$&$1.86\times10^{-6}$
&$1.80\times10^{-6}$\\
10000&$6.07\times10^{-7}$&$7.42\times10^{-7}$&$7.46\times10^{-7}$
&$7.15\times10^{-7}$\\
15625&$2.07\times10^{-7}$&$2.63\times10^{-7}$&$2.64\times10^{-7}$
&$2.51\times10^{-7}$\\
22500&$5.80\times10^{-8}$&$7.65\times10^{-8}$&$7.73\times10^{-8}$
&$7.28\times10^{-8}$\\
30625&$1.24\times10^{-8}$&$1.73\times10^{-8}$&$1.76\times10^{-8}$
&$1.63\times10^{-8}$\\
40000&$1.86\times10^{-9}$&$2.74\times10^{-9}$&$2.79\times10^{-9}$
&$2.56\times10^{-9}$\\
50625&$1.40\times10^{-10}$&$2.19\times10^{-10}$&$2.25\times10^{-10}$
&$2.01\times10^{-10}$\\
62500&$1.74\times10^{-12}$&$2.84\times10^{-12}$&$2.96\times10^{-12}$
&$2.61\times10^{-12}$\\
\hline
\end{tabular}
\caption{Decay rate for $\eta\rightarrow e^{+}e^{-}\ e^{+} e^{-}$,
normalized to $\Gamma (\eta\rightarrow\gamma\gamma)$.}
\label{tab:eeee}
\end{center}
\end{table}
%***********************************************************************
%

The difference between Form 2 and 3 is very small until the cut is 
$10000\ MeV^{2}$. According to Eq.~(\ref{eq:error}), the uncertainty 
here is
$\sigma\approx 10^{-8}$, so it is not possible to see the
difference here. For the highest cut the distinction is larger but the
experimental uncertainty is up to the size of the value itself. So, 
unfortunately, it seems
not to be possible to choose between form-factor 2 and 3 from decay 
experiments with 
this kind of precision. In the next section we will see if it is 
possible in collision experiments like 
$e^{+}e^{-}\rightarrow e^{+}e^{-}PS$, where $PS$ is $\pi^{0}$, $\eta$
or $\eta'$.
\clearpage
%***********************************************************************

% SECTION (Cross Sections)

\section{Cross Sections}
\label{sec-Cross}

%***********************************************************************

The cross section for a reaction is related to the probability for it
to happen. The name comes from the classical picture where the total 
cross section for e.g. $p\bar{p}$ scattering
is the area in which both particles have to be when they pass each other
for a reaction to happen. The cross section for e.g.
$p\bar{p}\rightarrow \pi^{+}\pi^{-}$ is then this area times the
probability that precisely $\pi^{+}\pi^{-}$ is in the final state and
nothing else. 

The process $e^{+}e^{-}\rightarrow PS\ e^{+}e^{-}$, where $PS$ is 
$\pi^{0}$, $\eta$ or $\eta'$, includes a meson-photon-photon vertex.
Therefore this is another opportunity to study the effects different 
form-factors will have. We will calculate the cross section
for all three processes with our four different form-factors to see
whether the difference will be larger here, possibly large enough to 
distinguish between Form 2 and 3.
 
Experimentally, these cross sections can be studied both at CESR, US, 
and at $\mrm{DA\Phi NE}$, Italy. At Cornell they are studied at CLEO 
with a center of mass energy $E_{CM}=10.58\ GeV$ and in Italy at KLOE 
with
 $E_{CM}=1.0192\ GeV$. We will look at both these energies in our 
calculations.

%***********************************************************************
\subsection{The Matrix Element}
%***********************************************************************

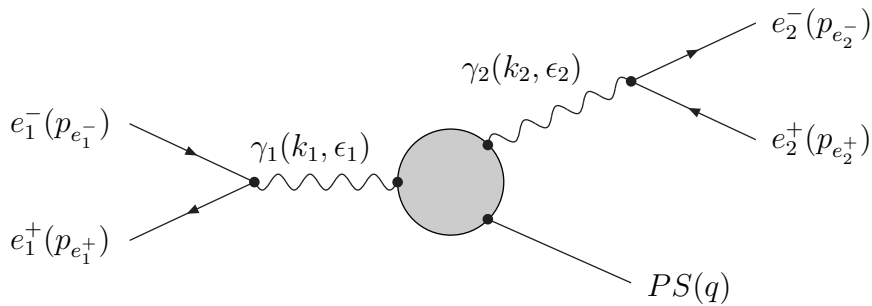
\begin{figure}[b] 
\begin{center}
\begin{picture}(400,150)
\GCirc(200,50){20}{0.8}
\Vertex(180,50){2}
\Vertex(214,64){2}
\Vertex(214,36){2}
\Vertex(268,88){2}
\Vertex(126,50){2}
\Photon(214,64)(268,88){3}{4.5}
\Line(214,36)(268,12)
\Photon(126,50)(180,50){-3}{4.5}
\ArrowLine(268,88)(315,110)
\ArrowLine(315,66)(268,88)
\ArrowLine(126,50)(79,28)
\ArrowLine(79,72)(126,50)
\put(322,63){$e^{+}_{2}(p_{e^{+}_{2}})$}                                
\put(322,107){$e^{-}_{2}(p_{e^{-}_{2}})$}
\put(34,25){$e^{+}_{1}(p_{e^{+}_{1}})$}
\put(34,69){$e^{-}_{1}(p_{e^{-}_{1}})$}
\put(275,7){$PS(q)$}
\put(125,60){$\gamma_{1}(k_{1},\epsilon_{1})$}
\put(205,90){$\gamma_{2}(k_{2},\epsilon_{2})$}
%\put(200,1){\textbf{(b)}}
\end{picture}
\end{center}
\caption{Diagram contributing to the cross section for 
$e^{+}e^{-}\rightarrow PS\ e^{+}e^{-}$.}
\label{fig:CSI}
\end{figure}

\begin{figure}[t] 
\begin{center}
\begin{picture}(400,170)
\GCirc(200,90){20}{0.8}
\Vertex(160,30){2}
\Vertex(160,150){2}
\Vertex(220,90){2}
\Vertex(186,104){2}
\Vertex(186,76){2}
\ArrowLine(160,30)(79,10)
\ArrowLine(315,10)(160,30)
\ArrowLine(79,170)(160,150)
\ArrowLine(160,150)(315,170)
\Photon(160,30)(186,76){3}{3.5}
\Photon(160,150)(186,104){-3}{3.5}
\Line(220,90)(280,90)
\put(322,3){$e^{+}_{2}(p_{e^{+}_{2}})$}                               
\put(322,167){$e^{-}_{2}(p_{e^{-}_{2}})$}
\put(34,3){$e^{+}_{1}(p_{e^{+}_{1}})$}
\put(34,167){$e^{-}_{1}(p_{e^{-}_{1}})$}
\put(287,87){$PS(q)$}
\put(180,127){$\gamma_{3}(k_{3},\epsilon_{3})$}
\put(180,50){$\gamma_{4}(k_{4},\epsilon_{4})$}
%\put(200,130){\textbf{(a)}}
\end{picture}
\end{center}
\caption{Second diagram contributing to the cross section for 
$e^{+}e^{-}\rightarrow PS\ e^{+}e^{-}$.}
\label{fig:CSII}
\end{figure}
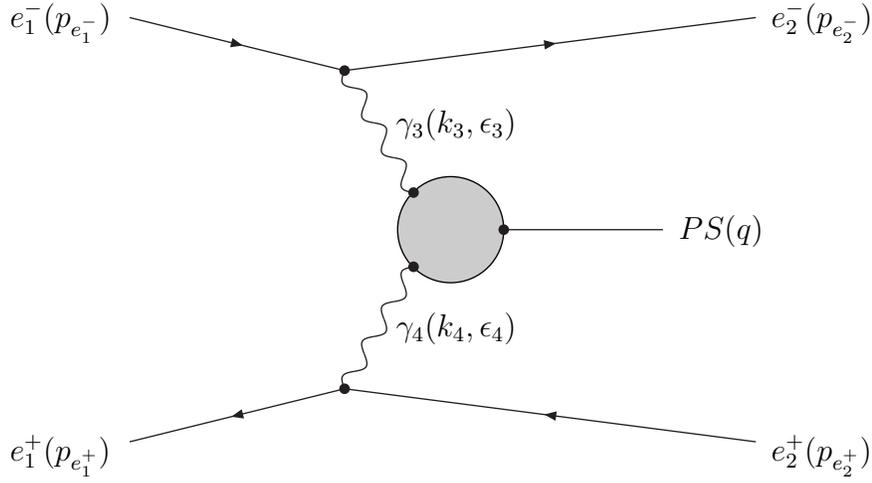

Irrespective of the pseudo-scalar created, two diagrams contribute
to the process, depicted in Fig.~\ref{fig:CSI} and 
Fig.~\ref{fig:CSII}.
%
%***********Fig~\ref{fig:} Cross Section I******************************
%
%***********************************************************************
%
%***********************************************************************
%
As seen, the process includes the same vertices and 
propagators as the earlier treated process 
$\eta\rightarrow e^{+}e^{-}\ e^{+}e^{-}$. Then, for reasons of crossing
symmetry, the matrix element should
also look the same, with a different numerical factor for 
$\pi^{0}$ and $\eta'$.
 But not exactly the same. The creation of one of the 
electron-positron pairs before corresponds to annihilation of a 
positron-electron pair now. This taken into account, the amplitude is
\begin{equation}
A(e^{+}e^{-}\rightarrow PS\ e^{+}e^{-})=-\frac{ie^{4}}{C
\pi^{2}\ F_{\pi}}\ \varepsilon^{\mu\nu\alpha\beta}
\ q_{\beta}\times[A_{1}-A_{2}]
\end{equation}
with
\begin{displaymath}
A_{1}=\frac{(\bar{\mrm{e}}_{1}\gamma_{\nu}
\mrm{e_{1}})k_{1\mu}(\bar{\mrm{e}}_{2}\gamma_{\alpha}
\mrm{e_{2}})+(\bar{\mrm{e}}_{1}\gamma_{\alpha}\mrm{e_{1}})k_{2\mu}
(\bar{\mrm{e}}_{2}\gamma_{\nu}\mrm{e_{2}})}{k_{1}^{2}k_{2}^{2}}
\end{displaymath}
\begin{displaymath} 
A_{2}=\frac{(\bar{\mrm{e}}_{1}\gamma_{\alpha}
\mrm{e_{2}})k_{3\mu}(\bar{\mrm{e}}_{2}\gamma_{\nu}
\mrm{e_{1}})+(\bar{\mrm{e}}_{1}\gamma_{\nu}\mrm{e_{2}})k_{4\mu}
(\bar{\mrm{e}}_{2}\gamma_{\alpha}\mrm{e_{1}})}{k_{3}^{2}k_{4}^{2}}  
\end{displaymath}
and
\begin{displaymath}
C= \left \{
\begin{array}{rcl}
8 & \mathit{for} & \pi^{0} \\ 
8\sqrt{3} & \mathit{for} & \eta \\
4\sqrt{3/2} & \mathit{for} & \eta' 
\end{array}
\right \}
\end{displaymath}
where the difference from the earlier case lies in which part of the 
electron and positron fields that are used. In the calculations we have
used an effective C for $\eta$ and $\eta'$, which reproduces the correct
decay rates for them, \mbox{$\Gamma(\eta\rightarrow \gamma \gamma)=
4.6\times10^{-4}\ MeV$} and 
\mbox{$\Gamma(\eta'\rightarrow \gamma \gamma)=4.27\times 10^{-3}\ MeV$}.
 
The amplitude is then squared and the result is the matrix element 
squared.

%***********************************************************************
\subsection{The Kinematics}
%***********************************************************************

The general formula for the cross section with three particles in the 
final state is \cite{ref:Journal}
\begin{equation}
d\sigma=\frac{(2\pi)^{4}\ |M|^{2}\ d\Phi_{3}}
{4\sqrt{(\mathbf{p_{1}}\cdot \mathbf{p_{2}})^{2}-m_{1}^{2}m_{2}^{2}}},
\end{equation} 
with
\begin{displaymath}
d\Phi_{3}=\delta^{4}(\mathbf{p_{1}}+\mathbf{p_{2}}-\mathbf{p_{3}}-
\mathbf{p_{4}}-\mathbf{p_{5}})\ 
\frac{d^{3}\vec{p}_{3}}{(2\pi)^{3}\ 2E_{3}}
\ 
\frac{d^{3}\vec{p}_{4}}{(2\pi)^{3}\ 2E_{4}}
\ 
\frac{d^{3}\vec{p}_{5}}{(2\pi)^{3}\ 2E_{5}}
\end{displaymath}
where $\mathbf{p_{1}}$ and $\mathbf{p_{2}}$ are the four momenta of
the incoming particles and $\mathbf{p_{3}}$, $\mathbf{p_{4}}$ and
$\mathbf{p_{5}}$ are the momenta of the outgoing particles. 
In this case $d\Phi_{3}$ can be rewritten as
\begin{equation}
d\Phi_{3}=\displaystyle{\frac{|\vec{p}_{e^{+}_{2}}^{\ast}|
d\Omega_{e^{+}_{2}}^{\ast}\
|\vec{p}_{e^{-}_{2}}|d\Omega_{e^{-}_{2}}\ dm_{PSe^{+}_{2}}^{2}}
{16(2\pi)^{9}\ m_{PSe^{+}_{2}}\ E_{CM}}},
\end{equation}
where $m_{PSe^{+}_{2}}=\sqrt{(q+p_{e^{+}_{2}})^{2}}$ and $\ast$ means 
that the quantity is calculated in the rest frame of $e_{2}^{+}$ and
the meson. 
Furthermore, since the lab system, where we want to calculate our
quantities, is the same as the $E_{CM}$ system we have
\begin{eqnarray}
\mathbf{p_{1}} & = & \left(\frac{E_{CM}}{2},0,0,
\sqrt{\frac{E_{CM}^{2}}{4}-m_{e}^{2}}\right)
\\ \nonumber
\mathbf{p_{2}} & = & \left(\frac{E_{CM}}{2},0,0,
-\sqrt{\frac{E_{CM}^{2}}{4}-m_{e}^{2}}\right),
\end{eqnarray}
which implies
\begin{equation}
\sqrt{(\mathbf{p_{1}}\cdot \mathbf{p_{2}})^{2}-m_{1}^{2}m_{2}^{2}}=
E_{CM}\ |\vec{p}_{in}|,
\end{equation}
where $p_{in}$ is either $p_{1}$ or $p_{2}$.

%***********************************************************************
\subsection{The Result}
%***********************************************************************

The z-axis is set by the incoming particles, but one can rotate freely
around it. This means that one of the angles in $d\Omega_{e^{-}_{2}}$
becomes just $2\pi$. Furthermore, a factor 1/4 is added since one has 
to take the average over incoming spins. Then there are four degrees 
of freedom left and the expression for the cross section is
\begin{equation}
d\sigma=\frac{|\vec{p}_{e^{+}_{2}}^{\ast}|\ |\vec{p}_{e^{-}_{2}}|\ 
|M|^{2}}{2^{8}\ (2\pi)^{4}\ m_{PSe^{+}_{2}}\ E_{CM}^{2}\ |\vec{p}_{in}|}
\ dm_{PSe^{+}_{2}}^{2}\ d\Omega_{e^{+}_{2}}^{\ast}\ d\cos{\theta_{2}}.
\end{equation}
The matrix element squared is made up of scalar products of the four 
vectors $p_{e^{+}_{1}}$, $p_{e^{+}_{2}}$, $p_{e^{-}_{1}}$ and 
$p_{e^{-}_{2}}$.
Therefore one has to figure out the how these vectors look in some 
common frame. This is done by expressing them in a frame where they
are known and then rotating and boosting them to the lab frame. 

Then everything is known and one can integrate to get the full cross
section. This is again done by using VEGAS integration on all four
degrees of freedom. Cuts are introduced, this time directly on the
angle between the outgoing electron and positron and the z-axis. The 
lowest cut is constructed so that the electron and positron precisely
go through the tracking system of the detectors.
  
The results, in pbarns, are presented in Table~\ref{tab:pi1} to 
Table~\ref{tab:eta'2}. The numerical errors are maximum one in the last
digit.

We have used the following masses
\begin{displaymath}
\begin{array}{lll}
m_{\pi^{0}}=134.98\ MeV & m_{\eta}=547.5\ MeV & m_{\eta'}=957.8\ MeV,
\end{array}
\end{displaymath}
%
%***************Table of Cross Section I********************************
\begin{table}[hb]
\begin{center}
\begin{tabular}{||c|r|r|r|r||} \hline
cut\ ($|\cos{\theta}|<$) & Form 1 & Form 2 & Form 3 & Form 4 \\ \hline
0.9&$2.41$&$2.09$&$1.94$&$1.63$ \\
0.8&$1.38$&$1.33$&$1.21$&$0.919$\\
0.6&$0.596$&$0.690$&$0.627$&$0.409$\\
0.4&$0.266$&$0.355$&$0.328$&$0.191$\\
\hline
\end{tabular}
\caption{Cross section for $e^{+}e^{-}\rightarrow \pi^{0}\ e^{+}e^{-}$ 
with $E_{CM}=1.019\ GeV$ and extra cut.}
\label{tab:pi1}
\end{center}
\end{table}
%***********************************************************************
%
\clearpage
%***************Table of Cross Section II*******************************
\begin{table}[ht]
\begin{center}
\begin{tabular}{||c|r|r|r|r||} \hline
cut\ ($|\cos{\theta}|<$) & Form 1 & Form 2 & Form 3 & Form 4 \\ \hline
0.93&$4.43$&$6.83\times10^{-4}$&$6.20\times10^{-3}$
&$3.56\times10^{-2}$ \\
0.8&$1.76$&$3.69\times10^{-5}$&$5.25\times10^{-4}$
&$2.05\times10^{-2}$\\
0.6&$0.762$&$1.23\times10^{-5}$&$8.58\times10^{-5}$
&$1.32\times10^{-2}$\\
0.4&$0.355$&$6.85\times10^{-6}$&$2.39\times10^{-5}$
&$8.04\times10^{-3}$\\
\hline
\end{tabular}
\caption{Cross section for $e^{+}e^{-}\rightarrow \pi^{0}\ e^{+}e^{-}$ 
with $E_{CM}=10.58\ GeV$ and extra cut.}
\label{tab:pi2}
\end{center}
\end{table}
%***********************************************************************
%\clearpage
%***************Table of Cross Section III******************************
\begin{table}[h]
\begin{center}
\begin{tabular}{||c|r|r|r|r||} \hline
cut\ ($|\cos{\theta}|<$) & Form 1 & Form 2 & Form 3 & Form 4 \\ \hline
0.9&$1.26$&$0.974$&$0.963$&$0.871$ \\
0.8&$0.694$&$0.557$&$0.548$&$0.457$\\
0.6&$0.291$&$0.260$&$0.255$&$0.186$\\
0.4&$0.123$&$0.126$&$0.123$&$0.0800$\\
\hline
\end{tabular}
\caption{Cross section for $e^{+}e^{-}\rightarrow \eta\ e^{+}e^{-}$ with
$E_{CM}=1.019\ GeV$}
\label{tab:eta1}
\end{center}
\end{table}
%***********************************************************************
%
%***************Table of Cross Section IV*******************************
\begin{table}[hb]
\begin{center}
\begin{tabular}{||c|r|r|r|r||} \hline
cut\ ($|\cos{\theta}|<$) & Form 1 & Form 2 & Form 3 & Form 4 \\ \hline
0.93&$3.90$&$6.06\times10^{-4}$&$5.48\times10^{-3}$
&$3.15\times10^{-2}$ \\
0.8&$1.55$&$3.27\times10^{-5}$&$4.65\times10^{-4}$
&$1.81\times10^{-2}$\\
0.6&$0.673$&$1.09\times10^{-5}$&$7.63\times10^{-5}$
&$1.17\times10^{-2}$\\
0.4&$0.314$&$6.05\times10^{-6}$&$2.12\times10^{-5}$
&$7.10\times10^{-3}$\\
\hline
\end{tabular}
\caption{Cross section for $e^{+}e^{-}\rightarrow \eta\ e^{+}e^{-}$ with
$E_{CM}=10.58\ GeV$ and extra cut.}
\label{tab:eta2}
\end{center}
\end{table}
%***********************************************************************
%
%***************Table of Cross Section V********************************
\begin{table}[t]
\begin{center}
\begin{tabular}{||c|r|r|r|r||} \hline
cut\ ($|\cos{\theta}|<$) & Form 1 & Form 2 & Form 3 & Form 4 \\ \hline
0.9&$0.205$&$0.183$&$0.183$&$0.187$ \\
0.8&$0.114$&$0.100$&$0.100$&$0.102$\\
0.6&$0.0457$&$0.0391$&$0.0392$&$0.0399$\\
0.4&$0.0172$&$0.0145$&$0.0146$&$0.0148$\\
\hline
\end{tabular}
\caption{Cross section for $e^{+}e^{-}\rightarrow \eta'\ e^{+}e^{-}$ 
with $E_{CM}=1.019\ GeV$}
\label{tab:eta'1}
\end{center}
\end{table}
%***********************************************************************
%
%***************Table of Cross Section VI*******************************
\begin{table}[ht]
\begin{center}
\begin{tabular}{||c|r|r|r|r||} \hline
cut\ ($|\cos{\theta}|<$) & Form 1 & Form 2 & Form 3 & Form 4 \\ \hline
0.93&$6.61$&$1.05\times10^{-3}$&$9.38\times10^{-3}$
&$5.38\times10^{-2}$ \\
0.8&$2.65$&$5.65\times10^{-5}$&$8.02\times10^{-4}$
&$3.09\times10^{-2}$\\
0.6&$1.15$&$1.87\times10^{-5}$&$1.32\times10^{-4}$
&$2.00\times10^{-2}$\\
0.4&$0.536$&$1.03\times10^{-5}$&$3.65\times10^{-5}$
&$1.21\times10^{-2}$\\
\hline
\end{tabular}
\caption{Cross section for $e^{+}e^{-}\rightarrow \eta'\ e^{+}e^{-}$ 
with $E_{CM}=10.58\ GeV$ and extra cut.}
\label{tab:eta'2}
\end{center}
\end{table}
%***********************************************************************

%***********************************************************************
\subsubsection{Additional Cuts}
%***********************************************************************

As stated earlier, some of the form-factors corresponds to a 
physical picture with intermediate mesons e.g. $\rho$. Therefore  
the mathematical expression of those factors are related to a meson 
propagator,
\begin{equation}
\frac{1}{k^{2}-M^{2}},
\end{equation}
where $k$ is the four momenta and $M$ the mass of the meson. This is
however only an approximation of the full propagator, which should 
include different higher order corrections, such as loops. This 
would have changed the propagator to
\begin{equation}
\frac{1}{k^{2}-M^{2}-\Sigma},
\end{equation}
where $\Sigma$ is a complex quantity related to the decay rate of the 
propagating meson. 

Using this approximation introduces problems when $k^{2}=M^{2}$,
which will happen in some cases with the energies above. In the case
with the high energy, $E_{CM}=10.58\ GeV$, the problem arises with all
three of the particles. This is solved by introducing a new cut. All
events in which $k_{1}^{2}$, $k_{2}^{2}$, $k_{3}^{2}$ or $k_{4}^{2}$
is in the range $(M_{\rho}-300\ MeV)^{2}\rightarrow 
(M_{\rho}+300\ MeV)^{2}$ are not counted in the cross section. The 
reason for choosing $300\ MeV$ is that it is twice the decay width of 
the $\rho$.
A good approximation for $\Sigma$ is $iM_{\rho}\Gamma_{\rho}$, which
means that when $q=M_{\rho}\pm 2\Gamma_{\rho}$, the absolute value of
the term we have neglected can at most be one fourth of $q^{2}$, and
therefore should not make a large difference.

The same thing happens with the low energy and the pion. But this time
it is not possible to use the same cut as last time, since all our
events lie in that region. We solve this by cutting in a more narrow
interval, $M_{\rho}\pm 200\ MeV$. Then the neglected term can be up
to half the size of the terms left but that will have to do. 

Experimentally, it is all right to make a more narrow cut in the low
energy case. A narrow cut means that a small error in the measurement
of the energy will make much difference, but this is compensated by
the fact that low energy means better resolution, experimentally.

%***********************************************************************
\subsubsection{Experimental Uncertainties}
%***********************************************************************

Once again it is interesting to calculate the uncertainties in the
experiments. The number of events is in this case given by the formula
\begin{equation}
N=L\ \sigma\ \Delta t,
\end{equation}
where $L$ is the luminosity of the experiment and $\Delta t$ the 
running time. In both experiments the standard luminosity should be
$L=5\times 10^{32}\ cm^{-2}\ s^{-1}$ and we will calculate with one
year of running time, $\Delta t=10^{7}\ s$. 

As mentioned before, a first estimate of the uncertainty is 
$\pm \sqrt{N}$, where $N$ is the number of events. From this one can 
get 
\begin{equation}
\sigma\pm \sqrt{\frac{\sigma}{L\ \Delta t}}  
\end{equation}
which is the result we use.
\clearpage

%***********************************************************************
\subsubsection{Conclusions}
%***********************************************************************

Just as in the decay rates it is possible to distinguish between Form 1 
and 4 in all
of these experiments. However, here the difference is large between
Form 2 and 3 in some cases as well. 

In the high energy experiment it should actually be possible to 
discern all 
four form-factors for all three particles. For e.g. $\eta$ with the cut
$|\cos{\theta}|<0.93$, the results with errors are
\begin{equation}
\begin{array}{lrl}
\mrm{Form\ 1} & (3.90\pm 0.03)& \\
\mrm{Form\ 2} & (6\pm 3)&\times 10^{-4} \\
\mrm{Form\ 3} & (5\pm 1)&\times 10^{-3} \\
\mrm{Form\ 4} & (3.2\pm 0.3)&\times 10^{-2} 
\end{array}
\end{equation}
In the low energy experiments it is more difficult. There it is only
in the case with the pion that it should be possible to distinguish
between all four. The results, with cut $|\cos{\theta}|<0.9$ are
\begin{equation}
\begin{array}{lr}
\mrm{Form\ 1} & (2.41\pm 0.02) \\
\mrm{Form\ 2} & (2.09\pm 0.02) \\
\mrm{Form\ 3} & (1.94\pm 0.02) \\
\mrm{Form\ 4} & (1.63\pm 0.02) 
\end{array}
\end{equation}
On the other hand, the difference is visible for all the four
different cuts in this case.

The above numbers are calculated with the luminosity mentioned above
and a typical year, which totals to an integrated luminosity of 
5000~$pb^{-1}$.
\clearpage

%***********************************************************************

% SECTION (The Muon Anomalous Magnetic Moment)

\section{The Muon Anomalous Magnetic Moment}
\label{sec-Muon}

%***********************************************************************

A magnetic moment is associated with a rotating charged body. 
Classically this magnetic moment $\mu$ is related to the angular 
momentum $l$ of the rotation through
\begin{equation}
\mu=\frac{q}{2m}\ l,
\end{equation}
where $q$ is the charge and $m$ is the mass of the body. But particles
also have intrinsic angular momentum, spin. This also gives rise to
a magnetic moment according to 
\begin{equation}
\mu_{s}=g_{s}\ \frac{q}{2m}\ s,
\end{equation}
where $s$ is the spin angular momentum vector and $q$ and $m$
is the charge and the mass of the particle. $g_{s}$, known as the 
gyromagnetic ratio, is just a constant which, according to Dirac theory,
is exactly equal to 2 for electrons and muons. 

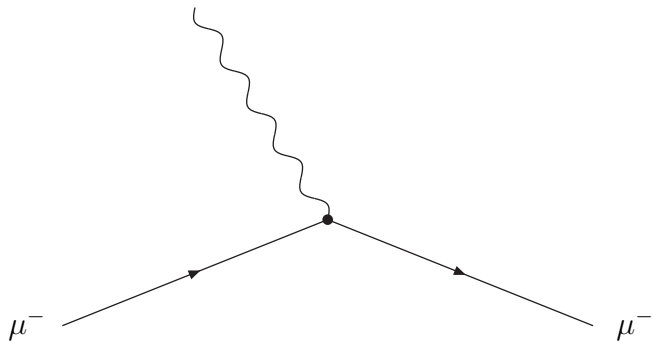
\begin{figure}[h] 
\begin{center}
\begin{picture}(400,150)
\Vertex(200,50){2}
\ArrowLine(100,10)(200,50)
\ArrowLine(200,50)(300,10)
\Photon(150,130)(200,50){-3}{5}
\put(80,7){$\mu^{-}$}
\put(310,7){$\mu^{-}$}
\end{picture}
\end{center}
\caption{Lowest order diagram for muon scattering by a magnetic field.}
\label{fig:MMM1}
\end{figure}

However, in Quantum Field Theory, which treats particles as quantized
fields as mentioned in the introduction, this is no longer true. 
Fig.~\ref{fig:MMM1} depicts the 
lowest order diagram for the interaction between a muon and a magnetic 
field, which gives $g_{s}=2$. 
%
%***********Figure of Magnetic Moment I*********************************
%***********************************************************************
%

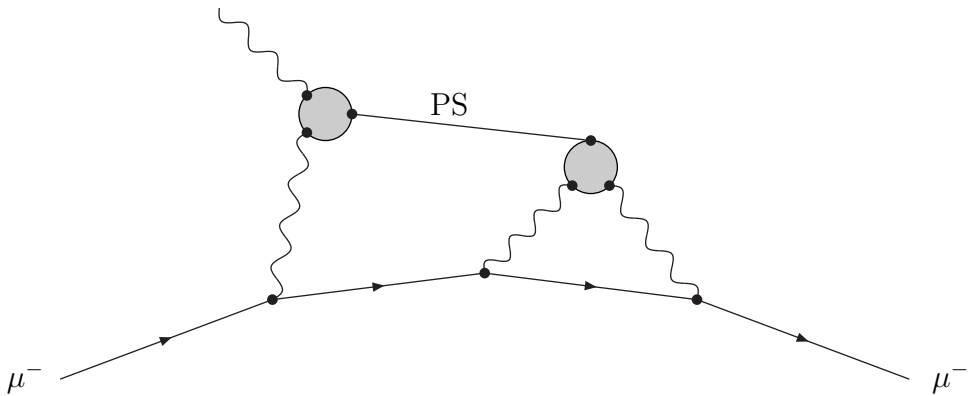
\begin{figure}[b] 
\begin{center}
\begin{picture}(400,150)
\Vertex(200,50){2}
\Vertex(120,40){2}
\Vertex(280,40){2}
\ArrowLine(40,10)(120,40)
\ArrowLine(120,40)(200,50)
\ArrowLine(200,50)(280,40)
\ArrowLine(280,40)(360,10)
\Photon(100,150)(133,117){-3}{3}
\put(20,7){$\mu^{-}$}
\put(370,7){$\mu^{-}$}
\GCirc(140,110){10}{0.8}
\Vertex(133,117){2}
\Vertex(133,103){2}
\Vertex(150,110){2}
\Photon(133,103)(120,40){-3}{4}
\GCirc(240,90){10}{0.8}
\Vertex(240,100){2}
\Vertex(233,83){2}
\Vertex(247,83){2}
\Line(150,110)(240,100)
\Photon(233,83)(200,50){-3}{3.5}
\Photon(247,83)(280,40){3}{3.5}
\put(180,110){PS}
\end{picture}
\end{center}
\caption{The dominant light-by-light contribution to the muon magnetic 
moment.}
\label{fig:MMM3}
\end{figure}

But then there are higher order corrections, which will change
the amplitude for the interaction between the muon and the magnetic 
field. This can be taken care of by changing the value of the 
gyromagnetic ratio by a small amount. To study these
corrections one defines the anomalous magnetic moment, $a_{\mu}$, as
\begin{equation}
a_{\mu}=\frac{1}{2}\ (g_{s}-2).
\end{equation} 
This will then be a measure of how large the higher order 
corrections from Quantum Field Theory are. 
At present the theoretical standard model estimate for $a_{\mu}$ is
\cite{ref:Hayakawa, ref:Bijnens, ref:Jegerlehner}
\begin{displaymath}
a_{\mu}(th)=11 659 178(7)\times 10^{-10},
\end{displaymath}
where the estimated error is in parentheses. $a_{\mu}$ gets 
contributions from five different types of diagrams.
\begin{enumerate}
\item Pure QED contributions are the largest corrections. They consist
e.g.\ of extra photon propagators or quark- and lepton loops, see 
Fig.~\ref{fig:MMM2}.
%
%***********Figure of Corrections I*************************************
\begin{figure}[t] 
\begin{center}
\begin{picture}(400,150)
\Vertex(100,50){2}
\Vertex(79,29){2}
\Vertex(121,29){2}
\ArrowLine(60,10)(79,29)
\ArrowLine(79,29)(100,50)
\ArrowLine(100,50)(121,29)
\ArrowLine(121,29)(140,10)
\Photon(100,130)(100,50){-3}{5}
\PhotonArc(100,50)(30,225,315){3}{3.5}
\put(40,7){$\mu^{-}$}
\put(150,7){$\mu^{-}$}

\Vertex(300,50){2}
\Vertex(300,75){2}
\Vertex(300,105){2}
\ArrowLine(260,10)(300,50)
\ArrowLine(300,50)(340,10)
\Photon(300,130)(300,105){-3}{2}
\Photon(300,75)(300,50){-3}{2}
%\Oval(300,90)(15,10)(0)
\ArrowArc(300,90)(15,90,-90)
\ArrowArc(300,90)(15,-90,90)
\put(240,7){$\mu^{-}$}
\put(350,7){$\mu^{-}$}

\end{picture}
\end{center}
\caption{Higher order QED diagrams for muon scattering by a magnetic 
field.}
\label{fig:MMM2}
\end{figure}
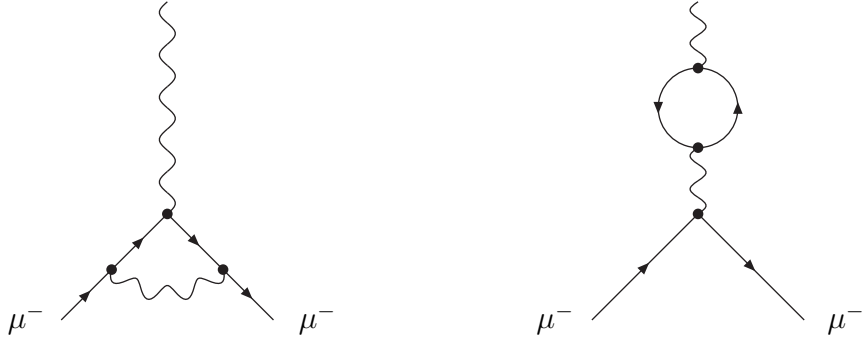
%***********************************************************************
%
\item Hadronic vacuum polarization, which means that all kinds of
hadron effects are added to the photons, e.g.\ quark-loops or 
intermediate mesons. 
  
\item Higher order hadronic vacuum polarization effects, which means 
even more hadron effects.

\item Hadronic light-by-light scattering contributions. These are
more complicated splittings and loops in the photon from the
magnetic field. The dominant diagram is depicted in Fig.~\ref{fig:MMM3}.

\item Electroweak contributions, which means that $Z^{0}$- and 
$W^{\pm}$-propagators are added in different ways.
\end{enumerate} 
%
%**********Figure of Light by light*************************************
%***********************************************************************
%
The piece that is studied in this paper is the one depicted in 
Fig.~\ref{fig:MMM3}, since it includes two meson-photon-photon vertices
and therefore also depends on the form-factors. We will calculate 
the contribution from this process to the anomalous magnetic moment for 
our form-factors 2,3 and 4 and compare the results.

A running experiment at Brookhaven National Laboratory measures 
the anomalous magnetic moment for the muon. In the future they hope to 
get an accuracy around \mbox{$\pm 4\times 10^{-10}$}, improving by a 
factor 
more than twenty the previous experimental determination at CERN. 
Therefore, it is important that the theoretical uncertainties are 
lowered to the 
same order as the aimed BNL uncertainty. Then the measurement can 
become a precision test of the quantum corrections of the electroweak
sector of the Standard Model. Even more, with the same low theoretical
and experimental uncertainties, and when combined with other high 
precision results, $a_{\mu}$ could become an excellent probe of 
physics beyond the Standard Model. 

%***********************************************************************
\subsection{The Result}
%***********************************************************************

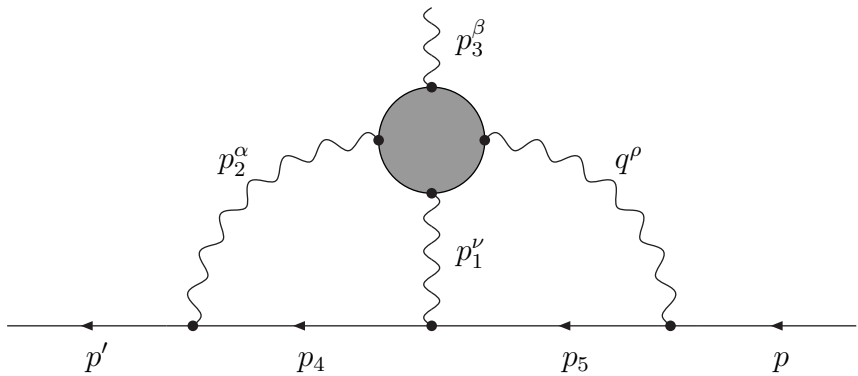
\begin{figure}[b] 
\begin{center}
\begin{picture}(400,150)
\GCirc(200,90){20}{0.6}
\Vertex(200,110){2}
\Vertex(200,70){2}
\Vertex(180,90){2}
\Vertex(220,90){2}
\Vertex(290,20){2}
\Vertex(200,20){2}
\Vertex(110,20){2}
\ArrowLine(360,20)(300,20)
\ArrowLine(300,20)(200,20)
\ArrowLine(200,20)(100,20)
\ArrowLine(100,20)(40,20)
\Photon(200,70)(200,20){3}{4}
\Photon(200,110)(200,140){3}{2.5}
\PhotonArc(180,20)(70,90,180){3}{7}
\PhotonArc(220,20)(70,0,90){-3}{7}
\put(330,5){$p$}
\put(250,5){$p_{5}$}
\put(150,5){$p_{4}$}
\put(70,5){$p'$}
\put(120,80){$p_{2}^{\alpha}$}
\put(270,80){$q^{\rho}$}
\put(210,45){$p_{1}^{\nu}$}
\put(210,125){$p_{3}^{\beta}$}
\end{picture}
\end{center}
\caption{Hadronic light-by-light contribution to $a_{\mu}$.}
\label{fig:LBL}
\end{figure}

All of the hadronic light-by-light contributions can be depicted as in 
Fig.~\ref{fig:LBL}, where the shaded blob includes all kinds of hadronic
effects. 
%
%************Figure of General LbL**************************************
%***********************************************************************
%
The numbering in the figure corresponds to one of six permutations, due
to the photons being identical particles. The amplitude for this 
process is \cite{ref:Bijnens}
\begin{eqnarray}
A & = & |e|^{7}\ A_{\beta}\ \int \frac{d^{4}p_{1}}{(2\pi)^{4}}\ 
\int \frac{d^{4}p_{2}}{(2\pi)^{4}}\ \frac{1}{q^{2}p_{1}^{2}p_{2}^{2}
(p_{4}^{2}-m^{2})(p_{5}^{2}-m^{2})}
\label{eq:MMM} 
\\ \nonumber \\ \nonumber
 & & \times \Pi^{\rho\nu\alpha\beta}(p_{1},p_{2},p_{3}) 
\bar{u}(p') \gamma_{\alpha} (\not p_{4}+m) \gamma_{\nu} 
(\not p_{5}+m) \gamma_{\rho}\ u(p)\\ \nonumber \\ \nonumber
 & & + \mrm{five\ more\ permutations,}
\end{eqnarray}
where $\Pi^{\rho\nu\alpha\beta}$, called the four-point function,
describes the shaded blob. In our case, $\Pi^{\rho\nu\alpha\beta}$
is just the product of the vertices and propagators corresponding
to Fig.~\ref{fig:MMM3}.  
For low momentum transfer, this amplitude can be expanded in a Taylor 
series in $(p-p')$. The linear term is then, 
with the momentum transfer set to zero,
by definition the anomalous magnetic moment. We calculate the 
pseudo-scalar light-by-light contribution to this term for form-factor 
2, 3 and 4. In 
doing that, the integrations in Eq.~(\ref{eq:MMM}) should go from zero 
to infinity. But it is interesting to see from which energy-regions
the contributions come. Therefore, we introduce a cut on the upper 
limit. This cut
is then raised until the result converges to some number, which
means that $a_{\mu}^{PS}$ has been saturated. 
The result is presented in Table~\ref{tab:MMM1} when $PS=\pi^{0}$, 
Table~\ref{tab:MMM2} when $PS=\eta$ and Table~\ref{tab:MMM3} when 
$PS=\eta'$. The numbers in parentheses are the errors quoted by VEGAS.
%
%***********Table of MMM I**********************************************
\begin{table}[t]
\begin{center}
\begin{tabular}{||c|r|r|r||} \hline
cut & Form 2 & Form 3 & Form 4 \\ 
($GeV$) & ($\times 10^{-10}$) & ($\times 10^{-10}$) & 
($\times 10^{-10}$) \\ \hline
0.4 & $-$2.683(2) & $-$2.707(2) & $-$2.821(2) \\
0.5& $-$3.440(3) & $-$3.449(3) & $-$3.683(3) \\
0.7& $-$4.448(5) & $-$4.622(5) & $-$4.936(5) \\
1.0& $-$5.152(7) & $-$5.538(8) & $-$5.972(8) \\
2.0& $-$5.60(1) & $-$6.43(1) & $-$6.96(1) \\
4.0& $-$5.64(1) & $-$6.67(2) & $-$7.21(2) \\
8.0& $-$5.63(1) & $-$6.74(2) & $-$7.27(2) \\
12.0& $-$5.63(1) & $-$6.70(2) & $-$7.25(2) \\
\hline
\end{tabular}
\caption{$a_{\mu}^{PS}$ when $PS=\pi^{0}$.}
\label{tab:MMM1}
\end{center}  
\end{table}
%***********************************************************************
%
%***********Table of MMM II*********************************************
\begin{table}[ht]
\begin{center}
\begin{tabular}{||c|r|r|r||} \hline
cut & Form 2 & Form 3 & Form 4 \\ 
($GeV$) & ($\times 10^{-10}$) & ($\times 10^{-10}$) & 
($\times 10^{-10}$) \\ \hline
0.4 & $-$0.4029(2) & $-$0.4071(2) & $-$0.4289(2) \\
0.5& $-$0.5819(3) & $-$0.5936(3) & $-$0.6353(3) \\
0.7& $-$0.8678(6) & $-$0.9088(6) & $-$0.9967(6) \\
1.0& $-$1.1058(9) & $-$1.211(1) & $-$1.357(1) \\
2.0& $-$1.281(1) & $-$1.545(2) & $-$1.768(2) \\
4.0& $-$1.301(2) & $-$1.647(3) & $-$1.900(3) \\
8.0& $-$1.302(2) & $-$1.670(3) & $-$1.937(3) \\
12.0& $-$1.301(2) & $-$1.676(4) & $-$1.943(3) \\
\hline
\end{tabular}
\caption{$a_{\mu}^{PS}$ when $PS=\eta$.}
\label{tab:MMM2}
\end{center}  
\end{table}
%***********************************************************************
%
The contributions from the three pseudo scalar mesons should then be 
added to get the final value of $a_{\mu}^{PS}$. The result is
%
%******Array of MMM PS**************************************************
\begin{equation}
\begin{array}{lc}
\mrm{Form\ Factor\ 2} & -7.94\times 10^{-10} \\
\mrm{Form\ Factor\ 3} & -9.73\times 10^{-10} \\
\mrm{Form\ Factor\ 4} & -10.85\times 10^{-10}
\end{array}
\end{equation}
%***********************************************************************
%
If nothing else were known of the form-factors, one would have to
take the mean value of all these and an uncertainty that covered
all of them. Then the final result would be
\begin{equation}
a_{\mu}^{PS}=-(9.51\pm 1.6)\times 10^{-10}.
\end{equation}
However, it looks like it is possible to exclude Form 4 already from
the experiment mentioned in Section 2.3. Therefore, we take a mean value
 of Form 2 and 3, which gives
\begin{equation}
a_{\mu}^{PS}=-(8.83\pm 0.9)\times 10^{-10}.
\end{equation}
This is our final result for $a_{\mu}^{PS}$ and since it isn't possible
to discern Form 2 and 3 yet, this must be the smallest possible
uncertainty on this value. This can be compared with the result in 
reference \cite{ref:Bijnens}
\begin{equation}
a_{\mu}^{PS}=-(8.5\pm 1.3)\times 10^{-10}
\end{equation}
and \cite{ref:Hayakawa}
\begin{equation}
a_{\mu}^{PS}=-(8.27\pm 0.64)\times 10^{-10}
\end{equation}
Even if neither Form 2 nor Form 3 satisfies
all the physical demands on a form-factor, our result means that one 
can not 
use arguments from existing experiments to lower the uncertainty more
than ours. To get better precision, experiments
like the cross sections mentioned in Section 4 have to be done, which
can better discern between different form-factors.
%
%***********Table of MMM III********************************************
\begin{table}[t]
\begin{center}
\begin{tabular}{||c|r|r|r||} \hline
cut & Form 2 & Form 3 & Form 4 \\ 
($GeV$) & ($\times 10^{-10}$) & ($\times 10^{-10}$) & 
($\times 10^{-10}$) \\ \hline
0.4 & $-$0.2595(1) & $-$0.2623(1) & $-$0.2770(1) \\
0.5& $-$0.3882(2) & $-$0.3964(2) & $-$0.4260(2) \\
0.7& $-$0.6113(4) & $-$0.6417(4) & $-$0.7102(4) \\
1.0& $-$0.8168(6) & $-$0.9008(6) & $-$1.0262(7) \\
2.0& $-$0.9874(9) & $-$1.219(1) & $-$1.443(1) \\
4.0& $-$1.008(1) & $-$1.324(2) & $-$1.597(2) \\
8.0& $-$1.010(1) & $-$1.351(2) & $-$1.645(2) \\
12.0& $-$1.009(1) & $-$1.353(2) & $-$1.654(3) \\
\hline
\end{tabular}
\caption{$a_{\mu}^{PS}$ when $PS=\eta'$.}
\label{tab:MMM3}
\end{center}  
\end{table}
%***********************************************************************
%

%***********************************************************************

% SECTION (Summary)

\section{Summary}
\label{sec-Sum}

%***********************************************************************

In this thesis we have been studying form-factors in meson-photon-photon
transitions. In this kind of transitions the exact form of the form
factors is not known and therefore different types are investigated.
We have studied four different factors, which all satisfy some, but
not all, of the constraints on form-factors coming from QCD.
They are 
% 
%************Array of form-factors**************************************
\\
\begin{displaymath}
\begin{array}{ll}
\mathit{\mrm{Form\ Factor\ 1:}} & F(k^{2}_{1},k^{2}_{2})=1 \\ 
\\
\mathit{\mrm{Form\ Factor\ 2:}} & F(k^{2}_{1},k^{2}_{2})=\displaystyle{
\frac{m_{\rho}^{4}}{(m_{\rho}^{2}-k_{1}^{2})(m_{\rho}^{2}-k_{2}^{2})}}\\
\\
\mathit{\mrm{Form\ Factor\ 3:}} & F(k^{2}_{1},k^{2}_{2})=\displaystyle{
\frac{m_{\rho}^{2}}{(m_{\rho}^{2}-k_{1}^{2}-k_{2}^{2})}} \\
\\
\mathit{\mrm{Form\ Factor\ 4:}} & F(k^{2}_{1},k^{2}_{2})=\displaystyle{
\frac{m_{\rho}^{4}-\frac{4\pi^{2}\ F_{\pi}^{2}}{N_{c}}\ (k_{1}^{2}+
k_{2}^{2})}{(m_{\rho}^{2}-k_{1}^{2})(m_{\rho}^{2}-k_{2}^{2})}} \\
\end{array}
\end{displaymath}
\\
%***********************************************************************
%

The idea is to calculate a set of physical quantities whose value 
depend on the choice of form-factor. This is done to
see if it should be possible to distinguish between the four in 
experiment.  
Already in section 2, results from an experiment done by the CLEO
collaboration were presented, which more or less rules out Form 1 and 4.
   
In section 3 the decay rates of processes involving 
$\eta\rightarrow \gamma \gamma$ are calculated. The results are compared
with the expected uncertainty from an upcoming experiment WASA at 
CELSIUS, Uppsala. Here again it seems like they will be able to exclude
Form 1 and 4 but not choose between Form 2 and 3.

In section 4 the cross section of three processes, which include a
meson-photon-photon transition, are calculated. The results are again
compared with estimated uncertainties, this time from experiments
at CESR, US and at $\mrm{DA\Phi NE}$, Italy. It turns out that in this
type of experiments it should be possible to make a more precise
measurement of form-factors and possibly discern all four.

Finally, in section 5, we make some comments about how the present
uncertainty around the choice of form-factor affects the value for
the muon anomalous magnetic moment. When examining the pseudo-scalar 
part of the light-by-light contribution to $a_{\mu}$, we can
estimate the lowest possible error in the theoretical value by 
looking at how much it differs when using Form 2 and 3. Since no 
present experiment can discern between the two, the uncertainty
must be at least large enough to cover both values. To get a better
precision, experiments have to be done, which can better discern
between different form-factors. 
This is important since the muon anomalous magnetic moment is one
of the most precisely measured quantities in physics and furthermore
can be a probe for physics beyond the Standard Model.

\subsection*{Acknowledgements}

First, I would like to thank my supervisor, Johan Bijnens, for 
introducing me to this subject and for patiently answering all my 
questions around it. I would also like to thank all the people at the 
department who have helped me during this time, and my family and 
friends for their support.   

\pagebreak

\end{document}